\documentclass[prd,aps,showpacs,nofootinbib,tightenlines]{revtex4}
\usepackage{mathrsfs}
\usepackage{amsmath}
\usepackage{amssymb}
\usepackage{epsfig}
\usepackage{graphicx}
\usepackage{booktabs}
\usepackage{multirow}
\usepackage{subfigure}
\usepackage[section]{placeins}
\usepackage{float}
\usepackage{slashed}
\usepackage{longtable}
\usepackage[colorlinks, citecolor=blue, linkcolor=red, urlcolor=blue]{hyperref}
\allowdisplaybreaks[4]
\begin{document}

\newcommand{\psl}{ p \hspace{-1.8truemm}/ }
\newcommand{\nsl}{ n \hspace{-2.2truemm}/ }
\newcommand{\vsl}{ v \hspace{-2.2truemm}/ }
\newcommand{\epsl}{\epsilon \hspace{-1.8truemm}/\,  }

\title{ Higher twist corrections to doubly-charmed baryonic $B$ decays}

\author{Zhou Rui }\email{jindui1127@126.com}
\author{Zhi-Tian Zou }\email{zouzt@ytu.edu.cn}
\author{Ying Li }\email{liying@ytu.edu.cn}
\affiliation{ Department of Physics, Yantai University, Yantai 264005, China}
\date{\today}
\begin{abstract}
Baryonic two-body $B$ mesons decays have been measured experimentally for some time, which provides an excellent ground for studying the QCD of baryonic $B$ decays. In this work, we investigate the two-body doubly-charmed baryonic $B$ decays in the perturbative QCD (PQCD) approach including higher twist contributions to the light-cone distribution amplitudes (LCDAs). The charmed baryon LCDAs are included up to the twist four and the effect of the subleading component of the $B$ meson LCDAs is studied for details. With the inclusion of these higher-power contributions, the PQCD results on rates can explain the current experimental data well. Moreover, we note that the SU(3) symmetry breaking is important in the concerned processes and play an essential role in understanding the measurements of $\mathcal{B}(\overline{B}^0\to \Lambda_c^+\overline \Lambda_c^-)$ and $ \mathcal{B}({B}^-\to\Xi_c^0\overline \Lambda_c^-)$. We also evaluate the angular asymmetries for the first time, which have neither been measured experimentally nor calculated theoretically. All these predictions can be tested in future.
\end{abstract}



\maketitle

\section{Introduction}
It is well known that exclusive $B$-decays offer an excellent laboratory for testing factorization, uncovering the nature of the QCD dynamics, revealing potential signals of $CP$ violation, and searching for the effects of the new physics beyond the Standard Model (SM). Compared with the mesonic $B$-decays, the exclusive baryonic decays of $B$ mesons have many unique signatures due to the half integral spins of the final state baryons, which provides fruitful physical observables in experiments. Theoretical studies of the baryonic $B$ decays are important for the understanding of the whole the $B$-physics sector. Particularly, the $B$ factories and LHCb experiments have found some new phenomenons, such as the  threshold enhancement~\cite{Belle:2002bro,Belle:2003pwf}, hierarchy of branching fractions for multi-body modes~\cite{Chistov:2016kae}, and angular correlation puzzle~\cite{Belle:2007oni, Belle:2005mke, Cheng:2006bn, Suzuki:2006nn}, which pose a great challenge to theorists.

In the SM, baryons can be produced only as baryon-antibaryon pairs due to the conservation of the baryon-number. Many light mesons decay to baryon pair are forbidden because of the insufficient phase space. However, $B$ meson is heavy enough to allow a baryon-antibaryon pair or even doubly charmed baryon pair production in the final state, providing a unique setting for shedding light on the baryon-production mechanisms. The experimental measurements for baryonic $B$ decays have a long history, including the asymmetric $e^+e^-$ collider experiments  BaBar and Belle~\cite{BaBar:2014omp} and the $pp$ collision LHCb~\cite{LHCb:2014nix,Barnard:2013hka}. Reviews of the experimental and theoretical status of exclusive baryonic $B$ decays could be referred to Refs~\cite{Huang:2021qld, Cheng:2006nm}. However, the theoretical description of non-leptonic two-body $B$ decays is notoriously complicated.

The dynamics of baryonic $B$ decay is complex and poorly understood theoretically. One of the common and conspicuous features for baryonic $B$ decay is the threshold enhancement \cite{Rosner:2003bm}. As conjectured in Ref.~\cite{Hou:2000bz}, the dibaryon mass spectra shows an strong enhancement near the threshold, the decay rates of two-body baryon-antibaryon final states are suppressed relative to rates of multi-body final states containing the same baryon-antibaryon pair. In this configuration the baryon-antibaryon pair prefers to be produced collinearly rather than back-to-back in baryonic $B$ decays~\cite{Hou:2000bz, Suzuki:2006nn, Cheng:2006bn}. Moreover, Belle Collaboration~\cite{Belle:2002gir, Belle:2005gtu}  also observe the hierarchy of branching ratios for two-body modes: the decays to two charmed baryons have rates larger than the single charmed baryonic processes and the latter have larger branching ratios than the charmless decays. For example, the measured branching ratios have the pattern~\cite{pdg2022}
\begin{eqnarray}
\mathcal{B}(B^-\to\overline {\Lambda}_c^-\Xi_c^0) (\sim10^{-3})\gg \mathcal{B}(\bar B^0\to\overline  {p}\Lambda_c^+) (\sim10^{-5})\gg
\mathcal{B}(\bar B^0\to\overline  {p}p) (\sim10^{-8}),
\end{eqnarray}
for two-body baryonic $B$ decays. Note that the Cabibbo-Kobayashi-Maskawa (CKM) matrix element and the phase space factor does not lead to such a clear hierarchy   of branching ratios, there must be a certain mechanism to enhance or suppress specific processes. Earlier calculations based on QCD sum rules~\cite{Chernyak:1990ag} or the diquark model~\cite{Ball:1990fw}  predict that the single charmed baryonic $B$ decays are comparable with the double ones,  shows great tension with the experiments. Phenomenologically, the preference could also be qualitatively comprehended in terms of the threshold enhancement  as a charmed baryon is heavier than a light baryon and the thresholds for baryon pair with more charmed baryons are closer to the $B$ meson mass. There have been several work describe the threshold peaking behavior  based on the Gluonic and fragmentation mechanisms~\cite{Rosner:2003bm}, Pole models~\cite{Suzuki:2006nn}, and Intermediate $X(1835)$ baryonium bound state~\cite{Datta:2003iy}. However, how to quantitatively evaluate their absolute decay rates and how to understand their distinct dynamics deserve more detailed studies, especially for  the QCD-inspired approaches.

In this work, we attempt to provide some quantitative insights into their dynamics, and concentrate primarily  on the two-body doubly charmed baryonic $B$ decays, which have attracted both experimental and theoretical attention. In the past few years, the LHCb and Belle collaborations had published the observations of $B^-\rightarrow\bar {\Lambda}_c^-\Xi_c^0$~\cite{Belle:2018kzz}, $\bar{B}^0\rightarrow\bar {\Lambda}_c^-\Xi_c^+$~\cite{Belle:2019bgi}, and $\bar{B}^0_{(s)}\rightarrow \Lambda_c^+ \bar\Lambda_c^-$~\cite{LHCb:2014scu}, with the branching ratios ranging from $10^{-6}$ to $10^{-3}$. Also, some final states involving excited $\Xi_c$ states, such as $B^-\rightarrow\bar {\Lambda}_c^-(\Xi_c^{'0},\Xi_c^0(2645,2790))$, were observed by Belle~\cite{Belle:2019pze}, which provide an experimental research platform to investigate the $\Xi_c$ excitation spectrum in the baryonic $B$ decays. These experimental results on $B$ meson baryonic decays are of a great importance to advance the understanding of the QCD. The key is to understand quantitatively the decay mechanism of the two-body decays. Some theoretical predictions for their branching fraction are available. In Refs.~\cite{Cheng:2005vd,Cheng:2009yz},  the authors assume that $\bar B\rightarrow \Xi_c \bar\Lambda_c $ is produced from a soft $q\bar q$ quark pair through the $\pi$ and $\sigma$ meson exchanges rather than the hard gluon exchanges, the predicted branching ratios are  at the level of $10^{-3}$. This magnitude is also supported by the estimation from final state interactions (FSIs)~\cite{Chen:2006fsa}. Very recently, two-body doubly charmfed baryonic decays  were investigated  within the framework of SU(3) flavor symmetry~\cite{Hsiao:2023mud}. With the determination of the SU(3) flavour parameters, they have predicted the branching ratios, which can be well approximated agrees with the experimental data. These efforts  may allow a better understanding of the threshold enhancement, and explain the hierarchy trend of the branching fractions for baryonic $B$ decays.

As the PQCD has achieved a preliminary success in the mesonic $B$ decays~\cite{Kurimoto:2001zj, Lu:2002ny,Ali:2007ff, Wang:2010ni, Rui:2021kbn, Chai:2022ptk} and $\Lambda_b$ baryon decays~\cite{prd59094014, prd61114002, cjp39328, prd65074030, prd74034026, prd80034011, 220204804, 220209181, prd106053005, 221015357, 230213785, Rui:2023fpp, Rui:2023fiz,Yu:2024cjd}, it is natural to extend this approach to evaluate the decay rates of baryonic decays of $B$ mesons, and the first attempt to calculate the $\bar B^0\rightarrow\bar {p}\Lambda_c^+$ decay has been performed in~\cite{He:2006vz}. Our emphasis in this work is to establish PQCD formalism for the two-body doubly charmed baryonic $B$ decays. Although the energy release in the concerned channels is small due to the two heavy charmed baryons in the final states, similar processes such as decays of $B$ mesons to two charmed mesons have been well studied in the PQCD approach~\cite{Li:2009xf,Rui:2012qq}, with the best agreement with experiments. We still assume that these processes are dominated by hard gluon exchanges. In this case, it is believed that some higher-power corrections may be numerically important. Here we will analyze these processes  by including the subleading $B$ meson LCDAs and charmed baryon LCDAs up to twist 4 level. It will be shown that the inclusion of these higher-twist contributions improves  the agreement of our results with data. Aside from the decay branching ratio, many asymmetry parameters are also predicted for the first time, which could be checked by future experiments.

The layout of this paper is as following. After presenting the effective hamiltonian,  kinematics, and the light-cone distribution amplitudes (LCDAs) of the initial and final states in Sec.~\ref{sec:framework}, we give a general PQCD formalism for the two-body doubly charmed baryonic $B$ decays. The numerical results for the  invariant  amplitudes, decay branching ratios, and various asymmetries are presented in  Sec.~\ref{sec:results}. Finally, Sec.~\ref{sec:sum} will be the conclusion of this work. Some details of the factorization formulas are displayed in Appendix~\ref{sec:for}.

\section{Theoretical framework}\label{sec:framework}

\begin{figure}[!htbh]
	\begin{center}
	    \vspace{0.01cm} \centerline{\epsfxsize=12cm \epsffile{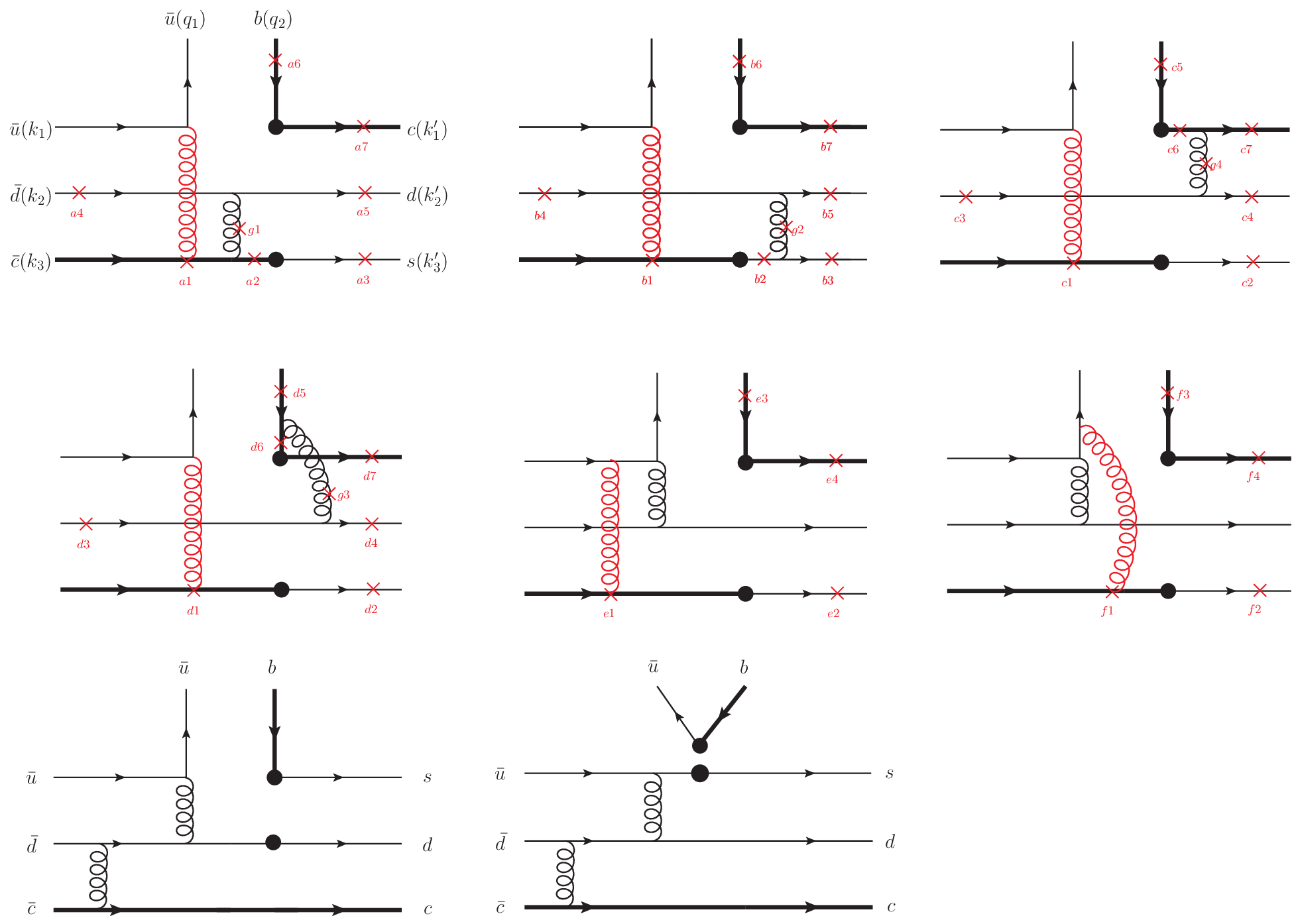}}
		\setlength{\abovecaptionskip}{1.0cm}
		\caption{Topological diagrams contributing to the ${B}^-\rightarrow\Xi_c^0\bar \Lambda_c^-$ decay,
where the solid black blob represents the vertex of the effective weak interaction. The heavy quarks $b$ and $c$ are shown in bold lines.The crosses indicate the possible connections of the gluon (red) attached to the spectator $\bar u$ quark. The first two rows are classified as the $W$-emission diagrams, while the last two topologies denote the penguin and weak annihilation diagrams, respectively. Here, we only draw one representative Feynman diagram for each of the last two topologies.}
		\label{fig:C}
	\end{center}
\end{figure}

The topology of the two-body baryonic $B$-meson decay is similar to that of the two-body baryon decay because they both involve two baryons and one meson in the initial and final states. In this case, the PQCD calculations of the two-body baryonic $B$-meson decay also start at the order of  $\mathcal{O}$$(\alpha_s^2)$, which requires at least two hard gluon exchanges in the leading order approximation. The quark diagrams for $ {B}^-\rightarrow\Xi_c^0\bar \Lambda_c^-$ mode involve the internal $W$-emission diagrams for $b \rightarrow c $ (the first two rows in Fig.~\ref{fig:C}), the penguin loop diagrams for $b \rightarrow s  $ transition (the first topology in the last row), and the weak annihilation diagrams for $b\rightarrow u$ (the last topology). Contributions from the penguin loop diagrams and the weak annihilation diagrams suffer highly loop suppressed and CKM suppression, respectively,  compared with the Cabibbo-favored $b \rightarrow c $ $W$-emission diagrams. Moreover, in both the penguin and weak annihilation diagrams, a back-to-back $c\bar c$ pair is created by a hard gluon, which is far off the mass shell. It should be harder than that produces a light quark pair in the $W$-emission diagrams, resulting in an additional dynamical suppression to the penguin and weak annihilation diagrams. An explicit evaluation shows that the contributions from the  penguin and weak annihilation topologies  can be safely neglected in the following analyses.

\begin{figure}[!htbh]
	\begin{center}
	    \vspace{0.01cm} \centerline{\epsfxsize=12cm \epsffile{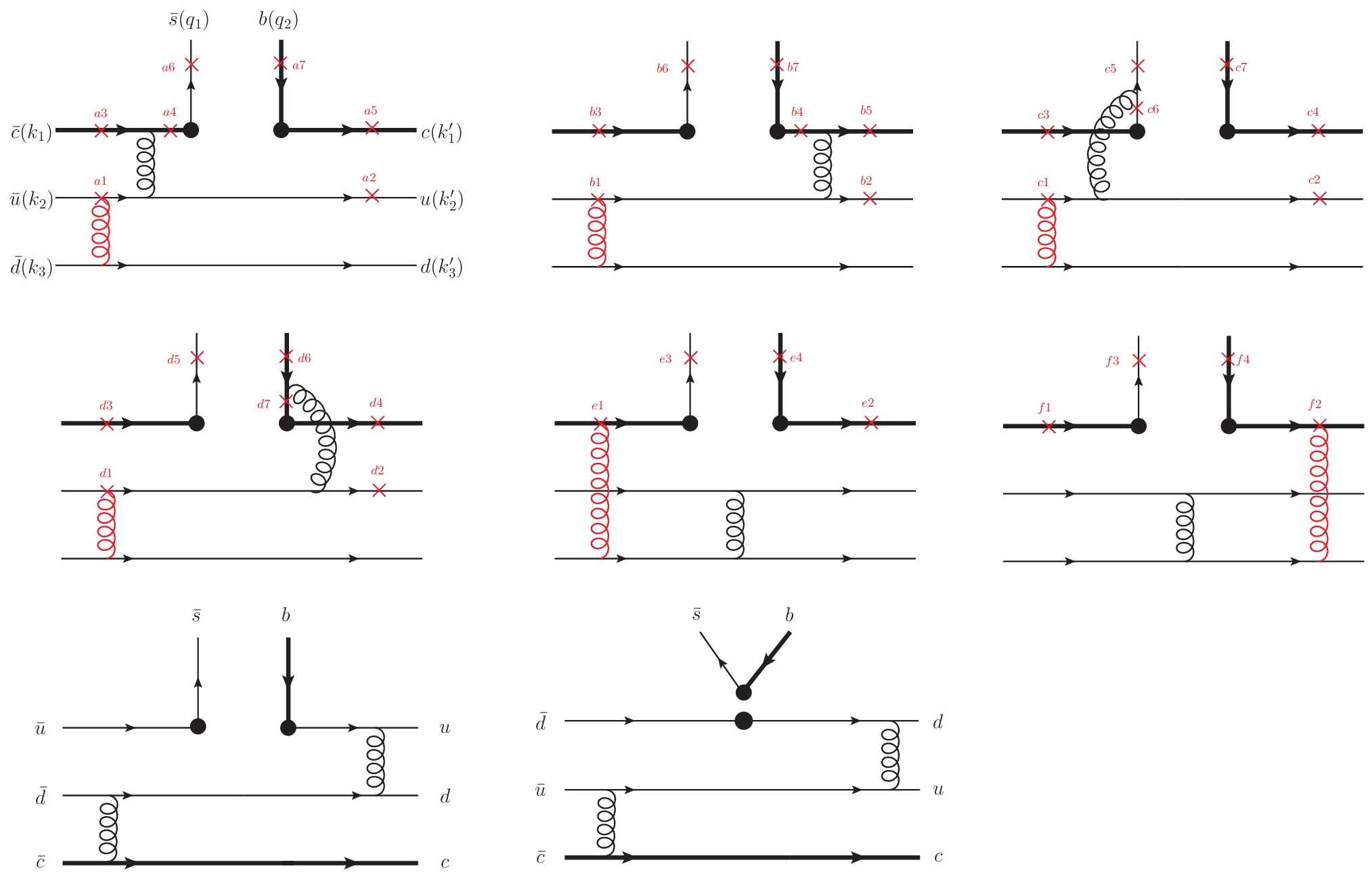}}
		\setlength{\abovecaptionskip}{1.0cm}
		\caption{Topological diagrams contributing to the $\bar {B}^0_s\rightarrow\Lambda_c^+\bar \Lambda_c^-$ decay.}
		\label{fig:E}
	\end{center}
\end{figure}

The $ \bar {B}^0 \rightarrow\Xi_c^+\bar \Lambda_c^-$  mode shares an isospin relation with $ {B}^-\rightarrow\Xi_c^0\bar \Lambda_c^-$, we do not repeat the corresponding diagrams here. The decay of  $ \bar {B}^0_s \rightarrow \Lambda_c^+\bar \Lambda_c^-$  is fed by the $W$-exchange and penguin-annihilation topologies as shown in Fig.~\ref{fig:E}. The first two rows are the dominant $W$-exchange diagrams induced by $b\rightarrow c$. In the last row, the first topology is another $W$-exchange diagram induced by $b\rightarrow u$ and the last one is the penguin-annihilation diagram induced by $b\rightarrow s$. Their contributions  suffer from severe suppression  as the same reason aforementioned. $\bar {B}^0 \rightarrow \Lambda_c^+\bar \Lambda_c^-$ proceeds through both the topological $W$-emission and $W$-exchange diagrams, whose diagrams can be obtained from Figs.~\ref{fig:C} and~\ref{fig:E}  with a suitable replacement of some quark flavors thus are not explicitly given here. All these topologies can be evaluated systematically in PQCD.

The kinematic variables of the initial and final states can be defined as follows. $p$ and $p^\prime$ denote the momenta of the antibaryon and baryon in the final state, respectively. $q=p+p^\prime$ denote the momentum of the $B$ meson. In the $B$ rest frame, the two final-state antibaryon and baryon move back to back with large momenta.The $B$ meson momentum in the light-cone coordinates  is written as $q=\frac{M}{\sqrt{2}}(1, 1, \textbf{0}_{T})$ with $M$ being the $B$ meson mass. We further assume that the antibaryon (baryon) is mainly along the plus (minus) direction and write
\begin{eqnarray}\label{eq:pq}
 p=\frac{M}{\sqrt{2}}(f^+,f^-,\textbf{0}_{T}),\quad p^\prime=\frac{M}{\sqrt{2}}(1-f^+,1-f^-,\textbf{0}_{T}).
\end{eqnarray}
The on-shell conditions of the baryon and antibaryon lead to
\begin{eqnarray}
f^\pm=\frac{1}{2}\left(1-r^2+\bar r^2
    \pm \sqrt{(1-r^2+\bar r^2)^2-4\bar r^2}\right),
\end{eqnarray}
where the mass ratio $r(\bar{r})=m(\bar m)/M$ with $m (\bar m)$ being the mass of (anti)baryon. The valence quark momenta inside the initial and final state hadrons, as shown in Fig.~\ref{fig:C}, are parametrized as
\begin{eqnarray}\label{eq:dyn}
q_1&=&\left(0,\frac{M}{\sqrt{2}}y,\textbf{q}_{T}\right),\quad q_2=q-q_1,\nonumber\\
k_1&=&\left(\frac{M}{\sqrt{2}}f^+x_1,0,\textbf{k}_{1T}\right),\quad
k_2=\left(\frac{M}{\sqrt{2}}f^+x_2,0,\textbf{k}_{2T}\right),\quad
k_3=p-k_1-k_2,\nonumber\\
k_1^\prime&=&p^\prime-k^\prime_1-k^\prime_2,\quad
k_2^\prime=\left(0,\frac{M}{\sqrt{2}}(1-f^-)x_2^\prime,\textbf{k}^\prime_{2T}\right),\quad
k_3^\prime=\left(0,\frac{M}{\sqrt{2}}(1-f^-)x_3^\prime,\textbf{k}^\prime_{3T}\right),
\end{eqnarray}
where $q_1$ is the spectator momentum  on the $B$ meson side.  $y$ and $\textbf{q}_T$ represent its longitudinal momentum fraction and transverse momentum, respectively. Here,  the bottom and charm quarks are considered to be massive, while other light quarks are massless. All the valence quarks carry transverse momenta $\textbf{k}^{(\prime)}_{lT}$ with $l=1,2,3$  to smear the end-point singularities. The parton longitudinal momentum fractions $x^{(\prime)}_l$ vary in the range of 0 to 1. For the $W$-exchange diagrams in Fig.~\ref{fig:E}, the parametrization of the kinematic variables in Eq.~(\ref{eq:dyn}) still apply except for $k_1$ and $k_3$ since in this case $k_1$  represents the anticharm quark momentum  instead of $k_3$. The new parametrization is chosen as $k_3=\left(\frac{M}{\sqrt{2}}f^+x_3,0,\textbf{k}_{3T}\right)$ and $k_1=p-k_2-k_3$.

In the course of the PQCD calculations, the necessary inputs contain the hadronic LCDAs of the initial and final states, which can be constructed via the nonlocal matrix elements. We next specify the relevant LCDAs for the present study. After the complete classification of three-quark LCDAs of the $\Lambda_b$ baryon in the heavy quark limit has been constructed ~\cite{plb665197}, the investigation of $\Lambda_b$ the heavy-baryon distribution amplitudes has made great progress in the last decade~\cite{plb665197,jhep112013191,epjc732302,plb738334,jhep022016179,Ali:2012zza}. However, the LCDAs of charmed baryons  have received less attention in the literature. Based on heavy-quark symmetry, we can use the same LCDAs for the baryon containing the charm quark and the bottom quark. Up to twist-4, the matrix element for the antitriplet of singly charmed baryons can be expressed in terms of the four LCDAs in the following way~\cite{Wang:2009hra, jhep112013191}:
\begin{eqnarray}\label{eq:LCDAs20}
\epsilon^{ijk}\langle0|q^i_{1\alpha}(t_1)q^j_{2\beta}(t_2)c^k_\gamma(0)|\mathcal{B}_c \rangle &=&\frac{ f^{(1)}}{8}
\Big[(\slashed{\bar n}\gamma_5C)_{\alpha\beta}\phi_2(t_1,t_2)+(\slashed{ n}\gamma_5C)_{\alpha\beta}\phi_4(t_1,t_2)\Big]u_\gamma \nonumber\\
&&+\frac{f^{(2)}}{4}\Big[(\gamma_5C)_{\alpha\beta}\phi_3^s(t_1,t_2)-\frac{i}{2}(\sigma_{\bar nn}\gamma_5C)_{\alpha\beta}\phi^{a}_3(t_1,t_2)\Big]u_\gamma
\end{eqnarray}
where $i$, $j$, and $k$ are color indices, and $\epsilon^{ijk}$ is the totally antisymmetric tensor. $\alpha,\beta,\gamma$ are the Dirac indices and $C$ is the charge conjugation matrix defined in terms of the Dirac matrices. $u $ denotes the on-shell Dirac spinor for the antitriplet $\mathcal{B}_c$. $n$ and $\bar n$ are two light-cone vectors satisfy $n^2=\bar n^2=0$ and  $n \cdot \bar n=2$. For the decay constants $f^{(1,2)}$,  we use the values $f^{(1,2)}_{\Lambda_c}=0.022 \pm 0.008$ GeV$^3$ and $f^{(1,2)}_{\Xi_c}=0.027\pm 0.008$ GeV$^3$ from the QCD sum rules~\cite{Wang:2010fq}. $\phi_{2,3,4}$ correspond twist-2, twist-3, and twist-4 LCDAs, respectively. In the limit of the exact SU(3) flavour symmetry, they have definite symmetry properties: $\phi_3^a$ is antisymmetric under the exchange $t_1\leftrightarrow t_2$, while others are symmetric. The corresponding model functions in the momentum space can be written as in the Gegenbauer expansion~\cite{Ali:2012zza,prd106053005}
 \begin{eqnarray}\label{eq:lcda}
\phi_2(x_2,x_3)&=&x_2x_3 m^4    \sum_{n=0}^2\frac{a_n}{{\varepsilon_n^4}}C_n^{3/2}\left(\frac{x_2-x_3}{x_2+x_3}\right)e^{-\frac{(x_2+x_3)m}{\varepsilon_n }},\nonumber\\
\phi_3^s(x_2,x_3)&=&(x_2+x_3)m^3  \sum_{n=0}^2\frac{a_n}{{\varepsilon_n^3}}C_n^{1/2}\left(\frac{x_2-x_3}{x_2+x_3}\right)
e^{-\frac{(x_2+x_3)m}{\varepsilon_n}},\nonumber\\
\phi_3^a(x_2,x_3)&=&(x_2+x_3)m^3 \sum_{n=1}^3\frac{b_n}{{\eta_n^3}}C_n^{1/2}\left(\frac{x_2-x_3}{x_2+x_3}\right)e^{-\frac{(x_2+x_3)m}{\eta_n }},\nonumber\\
\phi_4(x_2,x_3)&=&     m^2          \sum_{n=0}^2\frac{a_n}{{\varepsilon_n^2}}C_n^{1/2}\left(\frac{x_2-x_3}{x_2+x_3}\right)e^{-\frac{(x_2+x_3)m}{\varepsilon_n }},
\end{eqnarray}
with the Gegenbauer polynomials
\begin{eqnarray}
C_0^{\xi}(x)&=&1, \quad C_1^{\xi}(x)=2\xi x, \quad C_2^{\xi}(x)=2\xi(1+\xi)x^2-\xi. 
\end{eqnarray}
Note that the parameters $\varepsilon_{n}$, $ \eta_{n}$, $a_n$, and $b_n$ in Eq.~(\ref{eq:lcda})  differ  with distinct twists.
Their values evaluated at $\mu=1$ GeV have been determined in~\cite{Ali:2012zza}, which are collected in Table~\ref{tab:ac} for completeness.
The lowest-order Gegenbauer moments are set to 1 to comply with the normalizations~\cite{Ali:2012zza}.
We emphasize that though we employed the same forms and parameters as the bottom baryon LCDAs, but the mass $m$ in Eq.~(\ref{eq:lcda}) actually represents the charmed baryon mass.
This results in  the charmed baryon LCDAs behave differently, with exponentials decreasing more slowly in the large $x_{2,3}$ region than those in the bottom baryon LCDAs.
\begin{table}
\footnotesize
\caption{Values of the parameters $\varepsilon_{n}$  (GeV), $ \eta_{n}$ (GeV), $a_n$, and  $b_n$  appearing in the charmed baryon LCDAs.}
\label{tab:ac}
\begin{tabular}[t]{lcccccc}
\hline\hline
Baryon        &   &$\varepsilon_0$        & $\varepsilon_1$ & $\varepsilon_2$         & $a_1$ & $a_2$  \\ \hline
$     $      & $\phi_2$ &$0.201^{+0.143}_{-0.059}$ & $0$                & $0.551^{+\infty}_{-0.356}$ & $0$         & $0.391^{+0.279}_{-0.279}$\\
$\Lambda_c$  & $\phi_3^s$ &$0.232^{+0.047}_{-0.056}$ & $0$                & $0.055^{+0.010}_{-0.020}$  & $0$         & $-0.161^{+0.108}_{-0.207}$\\
$     $      & $\phi_4$ &$0.352^{+0.067}_{-0.083}$ & $0$                & $0.262^{+0.116}_{-0.132}$  & $0$         & $-0.541^{+0.173}_{-0.090}$\\ \hline
$     $      & $\phi_2$ &$0.228^{+0.068}_{-0.061}$ &$0.429^{+0.654}_{-0.281}$  &$0.449^{+\infty}_{-0.473}$ &$0.057^{+0.055}_{-0.034}$ & $0.449^{+0.236}_{-0.380}$\\
$\Xi_c$      & $\phi_3^s$ &$0.258^{+0.031}_{-0.038}$ &$0.750^{+0.308}_{-0.093}$  &$0.520^{+0.229}_{-0.060}$  &$0.339^{+0.261}_{-0.160}$ & $5.244^{+0.696}_{-1.132}$\\
$     $      & $\phi_4$ &$0.378^{+0.065}_{-0.080}$ &$2.291^{+\infty}_{-0.842}$ &$0.286^{+0.130}_{-0.150}$  &$0.039^{+0.030}_{-0.018}$ & $-0.090^{+0.037}_{-0.021}$\\

\hline\hline

        &  &$\eta_1$            & $\eta_2$     & $\eta_3$             & $b_2$ & $b_3$  \\ \hline
$\Lambda_c$  & $\phi_3^a$ &$0.324^{+0.054}_{-0.026}$ &$0$                &$0.633^{+0.0??}_{-0.0??}$  & $0$         &$-0.240^{+0.240}_{-0.147}$\\\hline
$\Xi_c$      & $\phi_3^a$ &$0.218^{+0.043}_{-0.047}$ &$0.877^{+0.820}_{-0.152}$  &$0.049^{+0.005}_{-0.012}$  &$0.037^{+0.032}_{-0.019}$ &$-0.027^{+0.016}_{-0.027}$\\

\hline\hline
\end{tabular}
\end{table}

For the baryonic $B$ decays, we need two  projectors for an outgoing baryon and an outgoing antibaryon. The former can be derived by
\begin{eqnarray}
\epsilon^{ijk}\langle \mathcal{B}_c |\bar {q}_{1\alpha}^i(t_1)\bar {q}_{2\beta}^j(t_2)\bar {c}_{\gamma}^k(0)|0\rangle&=&
-\gamma^0_{\alpha\alpha'}\gamma^0_{\beta\beta'}\gamma^0_{\gamma\gamma'}\epsilon^{ijk}\langle0|q^i_{1\alpha'}(t_1)q^j_{2\beta'}(t_2)c^k_{\gamma'}(0)|\mathcal{B}_c \rangle^{\dag}\nonumber\\&=&
\frac{ f^{(1)}}{8} \bar u_{\gamma}
[(C\gamma_5\slashed{\bar n})_{\beta\alpha}\phi_{2}(t_1,t_2)+(C\gamma_5\slashed{ n})_{\beta\alpha}\phi_4(t_1,t_2)] \nonumber\\
&&+\frac{f^{(2)}}{4}\bar u_{\gamma}[(C\gamma_5)_{\beta\alpha}\phi_3^s(t_1,t_2)+\frac{i}{2}(C\gamma_5\sigma_{\bar nn})_{\beta\alpha}\phi_3^a(t_1,t_2)].
\end{eqnarray}
Further perform the charge conjugation transform, we get the expression for the outgoing antibaryon
\begin{eqnarray}
\epsilon^{ijk}\langle \mathcal{\bar B}_c | {q}_{1\alpha}^i(t_1) {q}_{2\beta}^j(t_2) {c}_{\gamma}^k(0)|0\rangle&=&
-C_{\alpha'\alpha}C_{\beta\beta'}C_{\gamma\gamma'}\epsilon^{ijk}\langle \mathcal{B}_c |\bar {q}_{1\alpha'}^i(t_1)\bar {q}_{2\beta'}^j(t_2)\bar {c}_{\gamma'}^k(0)|0\rangle\nonumber\\&=&
-\frac{ f^{(1)}}{8} v_{\gamma}
[(\slashed{\bar n}\gamma_5 C)_{\beta\alpha}\phi_2(t_1,t_2)+(\slashed{ n}\gamma_5 C)_{\beta\alpha}\phi_4(t_1,t_2)] \nonumber\\
&&+\frac{f^{(2)}}{4}v_{\gamma}[(\gamma_5 C)_{\beta\alpha}\phi_3^s(t_1,t_2)+\frac{i}{2}(\sigma_{\bar nn}\gamma_5 C)_{\beta\alpha}\phi_3^a(t_1,t_2)].
\end{eqnarray}
where $v=C\bar {u}^T$ denote the spinor of an antiparticle.

As a heavy-light system,  the $B$-meson distribution amplitude in fact consists of two components~\cite{Grozin:1996pq,Beneke:2000ry}
\begin{eqnarray}\label{eq:bb}
\Phi_{B} =\frac{i}{\sqrt{2N_c}}(\rlap{/}{q}+M)\gamma_5\left(\frac{\rlap{/}{n_+}}{\sqrt{2}}\phi^+_{B} +\frac{\rlap{/}{n_-}}{\sqrt{2}}\phi^-_{B}\right),
\end{eqnarray}
where $\phi^{\pm}_{B}$ are two leading-twist distribution amplitudes with different asymptotic behaviors, which obey the normalizations
\begin{eqnarray}\label{eq:nou}
\int_0^1\phi^{\pm}_{B}(y)d y=\frac{f_{B}}{2\sqrt{2N_c}}.
\end{eqnarray}
By using the identity $(\rlap{/}{q}+M)\gamma_5(1+\frac{\rlap{/}{n_+}}{\sqrt{2}}+\frac{\rlap{/}{n_-}}{\sqrt{2}})=0$, Eq.(\ref{eq:bb}) can be written as
\begin{eqnarray}\label{eq:bbp}
\Phi_{B} =-\frac{i}{\sqrt{2N_c}}(\rlap{/}{q}+M)\gamma_5\left(\phi^-_{B} +\frac{\rlap{/}{n_+}}{\sqrt{2}}(\phi^-_{B}-\phi^+_{B} )\right),
\end{eqnarray}
which corresponds to the choice of $q_1$ as shown in Eq.~(\ref{eq:dyn})~\cite{Kurimoto:2006iv}. In this case,  $\phi_{B}=\phi^-_{B}$ is the leading component  and  $\bar \phi _{B}=\phi^-_{B}-\phi^+_{B}$  the subleading one. It has been pointed out that the  subleading component contribution is power suppressed,  thus its contribution can be neglected compared to that of leading one~\cite{Kurimoto:2001zj}. However, the recent updated studies on the nonleptonic $B$ decays show the subleading contribution are in fact comparable with those from the next-to-leading order corrections~\cite{Yang:2020xal,Yang:2022ebu}. In the following analysis we also consider this power-suppressed contribution and investigate its influence on baryonic $B$ decays. For numerical estimation, we use the conventional Gaussian model~\cite{Kurimoto:2001zj}
\begin{eqnarray}
\phi^-_{B}(y,b_q)=N_B y^2(1-y)^2\exp\left[-\frac{y^2M^2}{2\omega^2_b}-\frac{\omega^2_bb_q^2}{2}\right],
\end{eqnarray}
with $b_q$ being the impact parameter conjugate to the parton transverse momentum $q_{T}$. $N_B$ is the normalization constant which related to the $B$ meson decay constant $f_B$ via Eq.~(\ref{eq:nou}). We take the shape parameter $\omega_b=0.40$ GeV for $B_{u,d}$ mesons and $\omega_b=0.48$ GeV for a $B_s$ meson~\cite{Hua:2020usv}. $\phi^+_{B}$ can be obtained by solving the equation of motion $\phi^-_{B}=-y\frac{d\phi^+_{B}}{dy}$ without three-parton contributions~\cite{Grozin:1996pq}, whose explicit expression could be found in~\cite{Kurimoto:2006iv, Yang:2020xal}. For more alternative models of the $B$ meson DA and the  subleading contributions, one can refer to Refs.~\cite{prd102011502, prd70074030, Li:2012md, Li:2014xda, Li:2012nk,Han:2024min,Han:2024yun}.

For decays induced by $b\rightarrow q c \bar c$ transition, the related effective Hamiltonian is given by~\cite{Buchalla:1995vs}
 \begin{eqnarray}
\mathcal{H}_{eff}&=&\frac{G_F}{\sqrt{2}} \{V_{cb}V^*_{cq}[C_1(\mu)O_1(\mu)+C_2(\mu)O_2(\mu)]-\sum_{k=3}^{10}V_{tb}V^*_{tq}C_k(\mu)O_k(\mu)\}+\mathrm{H.c.},
\end{eqnarray}
with $q=d,s$. $G_F$ is the Fermi constant, and $V $ are the CKM matrix elements, $C_l(\mu)$ are Wilson coefficients  at the renormalization scale $\mu$ and the four quark operators $O_l$ are
\begin{eqnarray}
O_1&=& \bar{c}_i \gamma_\mu(1-\gamma_5) b_j  \otimes \bar{q}_j  \gamma^\mu(1-\gamma_5) c_i, \nonumber\\
O_2&=& \bar{c}_i \gamma_\mu(1-\gamma_5) b_i \otimes \bar{q}_j  \gamma^\mu(1-\gamma_5) c_j,  \nonumber\\
O_3&=& \bar{q}_i  \gamma_\mu(1-\gamma_5) b_i  \otimes \sum_{q'} \bar{q}'_j \gamma^\mu(1-\gamma_5) q'_j, \nonumber\\
O_4&=& \bar{q}_i  \gamma_\mu(1-\gamma_5) b_j \otimes \sum_{q'} \bar{q}'_j \gamma^\mu(1-\gamma_5) q'_i,  \nonumber\\
O_5&=& \bar{q}_i  \gamma_\mu(1-\gamma_5) b_i  \otimes \sum_{q'} \bar{q}'_j \gamma^\mu(1+\gamma_5) q'_j, \nonumber\\
O_6&=& \bar{q}_i  \gamma_\mu(1-\gamma_5) b_j \otimes \sum_{q'} \bar{q}'_j \gamma^\mu(1+\gamma_5) q'_i,  \nonumber\\
O_7&=&   \frac{3}{2}\bar{q}_i \gamma_\mu(1-\gamma_5) b_i  \otimes \sum_{q'} e_{q'}\bar{q}'_j \gamma^\mu(1+\gamma_5) q'_j, \nonumber\\
O_8&=&   \frac{3}{2}\bar{q}_i \gamma_\mu(1-\gamma_5) b_j \otimes \sum_{q'} e_{q'}\bar{q}'_j \gamma^\mu(1+\gamma_5) q'_i,  \nonumber\\
O_9&=&   \frac{3}{2}\bar{q}_i \gamma_\mu(1-\gamma_5) b_i  \otimes \sum_{q'} e_{q'}\bar{q}'_j \gamma^\mu(1-\gamma_5) q'_j, \nonumber\\
O_{10}&=&\frac{3}{2}\bar{q}_i \gamma_\mu(1-\gamma_5) b_j \otimes \sum_{q'} e_{q'}\bar{q}'_j \gamma^\mu(1-\gamma_5) q'_i.
\end{eqnarray}
The summation of $q'$ runs through $u, d, s, c$, and $b$ quarks. The decay amplitude of $B\rightarrow \mathcal{B}_c\mathcal{\bar B}_c$ can be described by sandwiching $\mathcal{H}_{eff}$  with the initial and final states,
\begin{eqnarray}\label{eq:amp}
\mathcal{M}=\langle \mathcal{B}_c\mathcal{\bar B}_c|\mathcal{H}_{eff}|B\rangle=\bar u [H_S+H_P\gamma_5]v,
\end{eqnarray}
where $H_S$ and $H_P$  are the respective $S$-wave and $P$-wave invariant amplitudes. They can symbolically be written as
\begin{eqnarray}\label{eq:ab}
H_{S/P}=\frac{16f_{\mathcal{ B}_c} f_{\mathcal{\bar B}_c} \pi^2 G_F}{27\sqrt{3}}\sum_{R_{ij}}
\int\mathcal{D}x\mathcal{D}b
\alpha_s^2(t_{R_{ij}})e^{-S_B-S_{\mathcal{B}_c}-S_{\mathcal{ \bar B}_c}}\Omega_{R_{ij}}(b,b',b_q)\sum_{\sigma=LL,SP}a_{R_{ij}}^{\sigma}H^{\sigma,S/P}_{R_{ij}}(x,x',y),
\end{eqnarray}
with the integration measures
\begin{eqnarray}
\mathcal{D}x&=&dx_1dx_2dx_3\delta(1-x_1-x_2-x_3)dx'_1dx'_2dx'_3\delta(1-x'_1-x'_2-x'_3)dy,\nonumber\\
\mathcal{D}b&=& d^2\textbf{b}_qd^2\textbf{b}_2d^2\textbf{b}_3d^2\textbf{b}'_2d^2\textbf{b}'_3.
\end{eqnarray}
The $\delta$ functions enforce momentum conservation.  $a_{R_{ij}}^{\sigma}$ denotes the product of the CKM matrix elements and the Wilson coefficients, and the labels $\sigma=LL$ and $SP$ refer to the contributions from $(V-A)(V-A)$ and $(S-P)(S+P)$ operators, respectively. $H^{\sigma}_{R_{ij}}$ is the numerator of the hard amplitude depending on the spin structure of final state. $\Omega_{R_{ij}}$ is the Fourier transformation of the denominator of the hard amplitude from the $k_T$ space to its conjugate $b$ space. The involving  variables $b$, $b'$ and $b_q$ conjugate to the parton transverse momenta $k_T$, $k'_T$ and $q_T$, respectively. The hard scale $t$ for each diagram is chosen as the maximal virtuality of internal particles including the factorization scales in a hard amplitude:
\begin{eqnarray}
t_{R_{ij}}=\max(\sqrt{|t_A|},\sqrt{|t_B|},\sqrt{|t_C|},\sqrt{|t_D|},w,w',1/b_q),
\end{eqnarray}
where $t_{A,B}$ are relevant to the two virtual quarks, while $t_{C,D}$ are associated with the two hard gluons. The infrared cut $w^{(')}$ is chosen to be the smallest inverse of a typical transverse distance among three valence quarks in baryon \cite{prd59094014}. Those quantities associated with specific diagram, such as $a_{R_{ij}}$, $H_{R_{ij}}$, and $t_{R_{ij}}$, are collected in Appendix.

The Sudakov exponents associated with  the initial and final states are written as~\cite{Ali:2007ff,prd80034011,Kundu:1998gv}
\begin{eqnarray}\label{eq:sud}
S_{B}&=&s(q^-_1,b_q)+\frac{5}{3}\int^t_{1/b_q}d \bar \mu \frac{\gamma(\alpha_s(\bar \mu))}{\bar \mu}\nonumber\\
S_{\mathcal{ B}_c}&=&s_c(k_1^{'-},cw')+\sum_{l=2}^3s(k_l^{'-},cw')+\frac{8}{3}\int^t_{cw'}d \bar \mu \frac{\gamma(\alpha_s(\bar \mu))}{\bar \mu}\nonumber\\
S_{\mathcal{ \bar B}_c}&=&s_c(k_1^{+},cw)+\sum_{l=2}^3s(k_l^{+},cw)+\frac{8}{3}\int^t_{cw}d \bar \mu \frac{\gamma(\alpha_s(\bar \mu))}{\bar \mu}
\end{eqnarray}
with the quark anomalous dimension $\gamma(\alpha_s(\bar \mu))=-\alpha_s(\bar \mu)/\pi$. Here we have included the Sudakov exponent $s_c$ associated with the charm quark, which carries large longitudinal momentum in the fast recoil region~\cite{prd61114002}. Its explicit form  with the inclusion of finite charm quark mass effects can be found in Ref~\cite{Liu:2023kxr}. The exponent $s$ for an energetic light quark is referred to Refs~\cite{prd59094014, Ali:2007ff}. The coefficient of the quark anomalous dimension in Eq.~(\ref{eq:sud}) is set to $5/3(8/3)$ for heavy meson (baryon) because the rescaled heavy-quark field adopted in the definition of the heavy hadron wave function has a self-energy correction different from that of the full heavy-quark field~\cite{Li:2004ja}. We assign a coefficient $c$ in front of $\omega^{(\prime)}$ to mimic theoretical uncertainties in resummation. Its value is expected to be of $\mathcal{O}(1)$ and is allowed to deviate slightly from unity. The best fit to the experimental data of $ \mathcal{B}({B}^-\rightarrow\Xi_c^0\bar \Lambda_c^-)$ determines the parameter as $c=1.05$, which is close to $c=1.14$ for the proton  one~\cite{Kundu:1998gv}.

\section{Numerical results}\label{sec:results}
We perform the numerical analysis in this section based on the factorization formulas derived above. The default values for input  parameters are given as follows:
\begin{itemize}
  \item Wolfenstein parameters~\cite{pdg2022}: $\lambda =0.22650$, \quad  $A=0.790$,  \quad $\bar{\rho}=0.141$, \quad $\bar{\eta}=0.357$.
  \item lifetimes (ps)~\cite{pdg2022}: $\tau_{B_s}=1.51$, \quad $\tau_{B_d}=1.52$, \quad $\tau_{B_u}=1.638$.
  \item Masses (GeV)~\cite{pdg2022}: $M_{B_{u,d}}=5.28$, \quad $M_{B_s}=5.37$, \quad $m_{\Lambda_c}=2.286$, \quad $m_{\Xi_c}=2.47$, \quad $m_b=4.8$, \quad  $m_c=1.275$.
  \item Decay constants~\cite{Wang:2010fq,He:2006vz,Rui:2021kbn}: $f_{B_{u,d}}=0.19$ GeV, \quad $f_{B_{s}}=0.24$ GeV, \quad $f_{\Lambda_c}=0.022$ GeV$^3$, \quad  $f_{\Xi_c}=0.027$ GeV$^3$.
\end{itemize}
Other nonperturbative parameters appearing in the hadron LCDAs have been specified in the preceding section.

\begin{table}[!htbh]
	\caption{Invariant amplitudes from the leading-twist ($\phi_B$) and subleading-twist ($\bar \phi_B$) $B$ meson LCDAs. The last column is their sum. }
	\label{tab:sub}
	\begin{tabular}[t]{lccccc}
	\hline\hline
Mode & Type &Amplitude &$\phi_B$ & $\bar \phi_B$ & $\phi_B+\bar \phi_B$ \\ \hline
& & $H_S$ & $1.2\times 10^{-7}+i8.3 \times 10^{-9}$ & $2.0\times 10^{-8}+i3.2 \times 10^{-8}$ & $1.4\times 10^{-7}+i4.0 \times 10^{-8}$ \\
$ {B}^-\rightarrow\Xi_c^0\bar \Lambda_c^-$ &$C$& $H_P$ & $-7.8\times 10^{-9}+i4.9 \times 10^{-8}$ & $-1.0\times 10^{-8}+i1.5 \times 10^{-8}$ & $-1.8\times 10^{-8}+i6.4 \times 10^{-8}$ \\
&&$ |\mathcal{M}|(\text{GeV})$ & $3.7 \times 10^{-7}$& $1.3 \times 10^{-7}$& $4.8 \times 10^{-7}$ \\
&&$H_S$ & $4.8\times 10^{-9}-i1.1 \times 10^{-8}$ & $5.0\times 10^{-9}+i8.6 \times 10^{-9}$ & $9.8\times 10^{-9}-i2.4 \times 10^{-9}$ \\
$\bar {B}^0_s\rightarrow\Lambda_c^+\bar \Lambda_c^- $ & $E$&$H_P$ & $-9.6\times 10^{-10}+i1.9 \times 10^{-8}$ & $5.8\times 10^{-9}-i3.0\times 10^{-9}$ & $4.8\times 10^{-9}+i1.6 \times 10^{-8}$ \\
&&$ |\mathcal{M}|(\text{GeV})$ & $1.1 \times 10^{-7}$& $4.5\times 10^{-8}$& $9.5 \times 10^{-8}$ \\
&& $H_S$& $7.8\times 10^{-9}+i6.1 \times 10^{-9}$ & $1.0\times 10^{-10}+i1.7 \times 10^{-9}$ & $7.9\times 10^{-9}+i7.8\times 10^{-9}$ \\
$ \bar {B}^0\rightarrow\Lambda_c^+\bar \Lambda_c^-$ &$C+E$&$H_P$ & $-2.5\times 10^{-9}+i1.5 \times 10^{-9}$ & $-2.8\times 10^{-9}+i2.3 \times 10^{-9}$ & $-5.3\times 10^{-9}+i3.8 \times 10^{-9}$ \\
&&$ |\mathcal{M}|(\text{GeV})$ & $3.0\times 10^{-8}$& $2.0 \times 10^{-8}$& $4.5\times 10^{-8}$ \\
		\hline\hline
	\end{tabular}
\end{table}	
We first compare the numerical results of the invariant amplitudes from the leading component ($\phi_B$) and subleading component ($\bar \phi_B$) of the $B$ meson LCDAs in Table~\ref{tab:sub}. The dominant topologies contributing to these decays are also indicated in the second column through the symbols $C$ (the internal $W$-emission) and $E$ (the $W$-exchange). It is observed that the subleading contributions are in fact comparable with the leading ones,  so the interference between the leading and subleading amplitudes has a significant impact on the  magnitude of amplitudes. Table~\ref{tab:sub} illustrates that the interference patterns for $C$ and $E$ amplitudes differ, with the former being constructive and the latter destructive. The $\bar {B}^0\rightarrow\Lambda_c^+\bar \Lambda_c^-$  decay amplitude receives contributions from $C$ and $E$ type topologies with the former as the main contribution. The resulting interference  is also constructive. It appears that the subleading contributions  from $\bar \phi_B$ can reach as much as $(30-70)\%$ of leading ones for the baryonic $B$ decays under consideration, which are larger than $30\%$ in the mesonic $B$ decays~\cite{Yang:2020xal, Yang:2022ebu, Lu:2002ny}. The inclusion of subleading correction can enhance  the total amplitudes $ |\mathcal{M}({B}^-\rightarrow\Xi_c^0\bar \Lambda_c^-)|$  and $ |\mathcal{M}(\bar {B}^0\rightarrow\Lambda_c^+\bar \Lambda_c^-)|$ by a factor of 1.3 and 1.5, respectively, and  lower $ |\mathcal{M}(\bar {B}^0_s \rightarrow\Lambda_c^+\bar \Lambda_c^-)|$ by a factor of 0.9 as shown in Table~\ref{tab:sub}. We will  see later that this subleading corrections improve the consistency between the PQCD predictions and the experimental data of the ${B}^-\rightarrow\Xi_c^0\bar \Lambda_c^-$ decays. We then conclude that the effect of the subleading component of the $B$ meson  distribution amplitude is indeed substantial for the concerned decays, which warns us that it is worth reexamining the doubly-charmed mesonic $B$ decays with the PQCD approach by considering the subleading component  $\bar \phi_B$ in the future.

\begin{table}[!htbh]
	\caption{Magnitude of amplitude $ |\mathcal{M}|(\text{GeV})$ from various twist combinations of the  baryon and antibaryon LCDAs. }
	\label{tab:twsit}
	\begin{tabular}[t]{lccc}
	\hline\hline
          & Twist-2                 & Twist-3                 & Twist-4                   \\ \hline
   $ {B}^-\rightarrow\Xi_c^0\bar \Lambda_c^-$  &&&\\
  Twist-2  & $3.5 \times 10^{-8}$    & $1.7\times 10^{-7}$    & $9.6\times 10^{-8}$      \\
  Twist-3  & $1.4 \times 10^{-7}$    & $1.9 \times 10^{-7}$    & $1.4 \times 10^{-7}$      \\
  Twist-4  & $1.1 \times 10^{-7}$    & $2.0 \times 10^{-7}$    & $1.6\times 10^{-7}$      \\
  $\bar {B}^0_s\rightarrow\Lambda_c^+\bar \Lambda_c^- $&&&\\
  Twist-2  & $3.2 \times 10^{-9}$    & $0$    & $1.5 \times 10^{-7}$     \\
  Twist-3  & $0$    & $1.5 \times 10^{-7}$    & $0$     \\
  Twist-4  & $5.8 \times 10^{-8}$    & $0$    & $1.5 \times 10^{-8}$      \\
  $ \bar {B}^0\rightarrow\Lambda_c^+\bar \Lambda_c^-$ &&&\\
  Twist-2  & $5.0 \times 10^{-9}$    & $2.6 \times 10^{-8}$    & $4.1 \times 10^{-8}$     \\
  Twist-3  & $2.1 \times 10^{-8}$    & $5.0 \times 10^{-8}$    & $1.5 \times 10^{-8}$      \\
  Twist-4  & $2.4 \times 10^{-8}$    & $3.0 \times 10^{-8}$    & $2.4 \times 10^{-8}$      \\
		\hline\hline
	\end{tabular}
\end{table}

We present the contributions to the amplitudes from various twist combinations of the baryon and antibaryon LCDAs in  Table.~\ref{tab:twsit}, in which both $ \phi_B$ and $\bar \phi_B$ contributions from $B$ meson LCDAs have been included. The twists in the rows correspond to the baryon LCDAs and those in the columns to the antibaryon ones. We learn from Table~\ref{tab:twsit} that the higher-twist baryon LCDAs give significant contributions to the decay amplitudes of all the three modes. For example, the contributions from the combination of twist-3 baryon and twist-3 antibaryon  LCDAs dominate the decay amplitudes, while the leading-twist contributions are generally an order of magnitude smaller. The main reason, as stated in Ref.~\cite{220204804}, is due to the  endpoint enhancement behaviors caused by the higher-twist baryon LCDAs, which could overcome the power suppressions from $r$ or $\bar r$ and remarkably increase higher-twist contributions.
Although the contributions of the twist-4-twist-4 combination are also important, they are still less than the dominant twist-3-twist-3 combination,
indicating the reliability of twist expansion of the baryon LCDAs.
As displayed in Table~\ref{tab:twsit}, the contributions from the combination of twist-3 baryon (antibaryon) and twist-2 or 4 antibaryon (baryon) LCDAs in $ \bar {B}^0_s \to \Lambda_c^+\bar \Lambda_c^-$ mode vanish. This observation can be understood easily as follows. According to the general definition of the non-local matrix elements in Eq.~(\ref{eq:LCDAs20}), the even (odd) twist LCDAs are always accompanied by even (odd) number of Dirac matrices. Therefore, the combination of baryon and antibaryon LCDAs with opposite parity always contains odd numbers of Dirac matrices, which must be exactly zero in a trace calculation. This situation appears  in the $W$-exchange diagrams, in which a contraction of light quark spinor indices leads to a trace involving both baryon and antibaryon LCDAs.

\begin{table}[!htbh]
	\caption{Theoretical predictions and experimental results on branching ratios  of $B\rightarrow \mathcal{B}_c\mathcal{\bar B}_c$  decays.
The label PQCD-I means the PQCD results without the  subleading correction from the $B$ meson LCDAs, and PQCD-II means the inclusions of the  subleading correction. The errors for these entries correspond to the uncertainties in the $B$ meson LCDAs, charmed baryon LCDAs, the scale dependence, and the Sudakov resummation, respectively. The model predictions by SU(3)~\cite{Hsiao:2023mud}  are also displayed here for comparison. The last column contains measured branching ratios or upper limits from the Particle Data Group (PDG)~\cite{pdg2022} or the original experimental literature~\cite{Belle:2018kzz, Belle:2019bgi, LHCb:2014scu}.}
	\label{tab:branching}
	\begin{tabular}[t]{lccccc}
	\hline\hline
Mode &Transition  & PQCD-I & PQCD-II  & SU(3)~\cite{Hsiao:2023mud} & Data~\cite{pdg2022,Belle:2018kzz,Belle:2019bgi,LHCb:2014scu} \\ \hline
$ {B}^-\rightarrow\Xi_c^0\bar \Lambda_c^-$ &$b\rightarrow sc\bar c$  &$(5.7^{+2.3+1.9+1.1+0.7}_{-1.6-2.5-0.9-0.6})\times10^{-4} $
&$(9.5^{+3.0+2.6+1.7+1.2}_{-2.3-3.5-1.4-1.1})\times10^{-4} $  &$7.8^{+2.3}_{-2.0}\times10^{-4}$ &$(9.5\pm2.3)\times10^{-4}$ \\
$\bar {B}^0\rightarrow\Xi_c^+\bar \Lambda_c^-$ &$b\rightarrow sc\bar c$ &$(5.3^{+2.1+1.7+1.0+0.6}_{-1.4-2.2-0.8-0.6})\times10^{-4} $
& $(8.8^{+2.7+2.6+1.5+1.1}_{-2.1-3.1-1.2-1.0})\times10^{-4} $    &$7.2^{+2.1}_{-1.9}\times10^{-4}$ &$(12\pm8)\times10^{-4}$ \\
$\bar {B}^0_s\rightarrow\Lambda_c^+\bar \Lambda_c^-$ &$b\rightarrow sc\bar c$  &$(5.0^{+0.7+0.3+1.4+1.4}_{-0.6-0.5-0.9-1.0})\times10^{-5} $
& $(4.0^{+0.7+0.2+0.9+1.0}_{-0.3-0.1-0.7-0.8})\times10^{-5} $     &$8.1^{+1.7}_{-1.5}\times10^{-5}$ &$  <9.9\times10^{-5}$ \\
$\bar {B}^0\rightarrow\Lambda_c^+\bar \Lambda_c^-$ &$b\rightarrow dc\bar c$  &$(4.0^{+2.5+1.7+0.7+0.6}_{-1.4-1.3-0.6-0.4})\times10^{-6} $
& $(8.8^{+4.4+3.5+1.1+1.0}_{-2.8-3.6-0.9-0.6})\times10^{-6} $        &$2.1^{+1.0}_{-0.8}\times10^{-5}$ &$  <1.6\times10^{-5}$ \\
		\hline\hline
	\end{tabular}
\end{table}	

Using above numerical result of  the decay amplitudes, it is straightforward to calculate the  branching ratio, reads
\begin{eqnarray}\label{eq:two}
\mathcal{B}&=&\frac{P_c\tau_B}{8\pi M^2}|\mathcal{M}|^2=\frac{P_c\tau_B}{8\pi M^2}(|H_S|^2Q_++|H_P|^2Q_-),
\end{eqnarray}
with $Q_{\pm}=M^2-(m\pm\bar m)^2$ and the spatial momentum of baryon $P_c=\sqrt{Q_+Q_-}/(2M)$. The corresponding numerical results are summarized in Table~\ref{tab:branching}. The results without (with) the subleading contribution, labeled by PQCD-I(II),  are also shown for comparison. Four types of theoretical uncertainties are estimated here. The first error is due to  the shape parameter $\omega_b$ in the $B$ meson distribution amplitudes. We consider the variation of its shape parameter within $10\%$ around its central value. The resulting uncertainty for the branching ratios amount to $20\%-50\%$. The second error bar includes the uncertainties due to variation of the Gegenbauer moments in the charmed baryon LCDAs as displayed in Table.~\ref{tab:ac}, added in quadrature. Numerical results demonstrate that the branching ratios are more sensitive to $a_2^{(3)}$ among these nonperturbative parameters. The third one  reflects the unknown next-to-leading-order QCD corrections characterized by the hard scale $t$, which is varied from $0.8t$ to $1.2t$. The last  uncertainty  is due to the parameter $c=1.05\pm0.05$ in the Sudakov resummation for the charmed baryon wave function. The variation of $c$ denotes the different partitions of radiative corrections into the perturbative Sudakov factor and the nonperturbative wave function~\cite{prd59094014}.
It is concluded that the absolute branching ratios suffer large theoretical uncertainties from the nonperturbative hadronic parameters in the present situation.

Predictions in the SU(3) approach~\cite{Hsiao:2023mud} and the data from the Particle Data Group~\cite{pdg2022} are also collected in Table~\ref{tab:branching} to make a comparison. On the experimental side, Belle collaboration has published the measurements of $ \mathcal{B}({B}^-\rightarrow\Xi_c^0\bar \Lambda_c^-)=(9.51\pm2.10\pm0.88)\times10^{-4}$~\cite{Belle:2018kzz} and $\mathcal{B}(\bar {B}^0\rightarrow\Xi_c^+\bar \Lambda_c^-)=(1.16\pm0.42\pm0.15)\times10^{-3}$~\cite{Belle:2019bgi}, where the first uncertainty is statistical and the second is systematic. The $95\%$ credibility level upper limits on $\bar{B}^0\rightarrow \Lambda_c^+ \bar\Lambda_c^-$ and $\bar{B}^0_s\rightarrow \Lambda_c^+ \bar\Lambda_c^-$ were determined to be $1.6\times10^{-5}$  and $9.9\times10^{-5}$ by LHCb~\cite{LHCb:2014scu}, respectively. However, the previous measurement performed by Belle yields a larger value of $\mathcal{B}(\bar{B}^0\rightarrow \Lambda_c^+ \bar\Lambda_c^-)=(2.2^{+2.2}_{-1.6}\pm 1.3)\times 10^{-5}$~\cite{Belle:2007lyc} with an upper limit of $6.2\times10^{-5}$ at $90\%$ confidence level. In the SU(3) approach~\cite{Hsiao:2023mud}, the $W$-emission and $W$-exchange amplitudes are determined using above available experimental data, which can be applied to explore  other channels based on the SU(3) flavour symmetry.  As indicated in Table~\ref{tab:branching},  the PQCD predictions without the subleading corrections are generally smaller than the SU(3) results and the data. As aforementioned, the interference patterns between the leading and subleading contributions from the $B$ meson LCDAs are different for the $C$ and $E$ amplitudes. After including the subleading corrections, the predicted branching ratios of $ {B}^-(\bar {B}^0)\rightarrow\Xi_c^0(\Xi_c^+)\bar \Lambda_c^-$ and $\bar {B}^0_s\rightarrow\Lambda_c^+\bar \Lambda_c^-$ will increases and decreases, respectively. Roughly speaking, the inclusion of the subleading corrections could alleviate the discrepancy and improve the agreement with the current data. In particular, our prediction on the branching ratio of ${B}^-\rightarrow\Xi_c^0\bar \Lambda_c^-$  is compatible with the world average value.
It is worth mentioning that the tuning of $c$ is not crucial for explaining the data because of the insensitivity of the predictions to $c$ shown in Table~\ref{tab:branching}.
The PQCD predictions for the $ \mathcal{B}(\bar {B}^0\rightarrow\Lambda_c^+\bar \Lambda_c^-)$ and $ \mathcal{B}(\bar {B}^0_s \rightarrow \Lambda_c^+\bar \Lambda_c^-)$ branching ratios reach half of the measured upper limits. They may be just around the corner.
Confronting with the results of SU(3)  approach \cite{Hsiao:2023mud}, our branching ratios for the first two modes are larger, however those of the last two processes are smaller. Nevertheless, considering the uncertainties of the two theoretical predictions, the disparities are not serious. An earlier calculation in Ref.~\cite{Cheng:2005vd} gave $\mathcal{B} ({B}^-\rightarrow\Xi_c^0\bar \Lambda_c^-)=(2.2^{+0.6+5.1+6.1}_{-0.6-1.9-1.9})\times10^{-3}$ and $ \mathcal{B}(\bar {B}^0\to\Xi_c^+\bar \Lambda_c^-)=(2.0^{+0.5+4.7+5.6}_{-0.6-1.7-1.7})\times10^{-3}$, which are considerably larger. Although the subsequent updated branching ratios for the two modes  are consistent with ours, their central value of $\mathcal{B}(\bar {B}^0\rightarrow\Lambda_c^+\bar \Lambda_c^-)=(5.2^{+2.3+0.6+2.6+0.0}_{-1.1-0.3-1.5-0.0})\times10^{-5}$~\cite{Cheng:2009yz} exceeds the current existing upper limit by a factor of 3.

\begin{table}[!htbh]
	\caption{Contributions to the invariant amplitudes of the $\bar {B}^0\rightarrow\Lambda_c^+\bar \Lambda_c^-$ decay from $C$ and $E$ topologies.
 The last two columns are their relative magnitude and phase, respectively.}
	\label{tab:CE}
	\begin{tabular}[t]{lcccc}
	\hline\hline
Amplitude  &$C$ & $E$ &$|\frac{E}{C}|$ &$\text{Arg}(E/C)$\\ \hline
$H_S$ &$1.0\times10^{-8}+i7.5\times10^{-9}$ &$-2.3\times10^{-9}+i3.0\times10^{-10}$ &0.18&2.37\\
$H_P$ &$-4.1\times10^{-9}+i7.4\times10^{-9}$ &$-1.1\times10^{-9}-i3.7\times10^{-9}$ &0.44&2.34\\
		\hline\hline
	\end{tabular}
\end{table}

The decay $\overline {B}^0\rightarrow\Lambda_c^+\bar \Lambda_c^-$  is of particular interest as it receives both the $W$-emission and $W$-exchange contributions. In the SU(3) approach,  the magnitudes of the two topological amplitudes were determined to be $|C|=1.29\pm0.18$ GeV$^3$ and $|E|=0.20\pm0.02$ GeV$^3$~\cite{Hsiao:2023mud}. The resultant ratio of $|2E/C|=0.31$  indicates the $W$-exchange contribution is important and nonnegligible. As pointed out in~\cite{Hsiao:2023mud}, the accuracy of the  current available data is not high enough to determine the $W$-exchange amplitude and the relative phase between the $W$-emission and $W$-exchange amplitudes simultaneously. To accommodate the experimental data, the relative phase was chosen as $\pi$ to maximize the destructive interference between the $W$-emission and $W$-exchange amplitudes. The PQCD predictions on the central values of the invariant amplitudes from the $W$-emission and $W$-exchange diagrams are presented separately in Table~\ref{tab:CE}, where the last two columns are the relative magnitude and phase, respectively. It should be noted that the topological amplitudes defined in~\cite{Hsiao:2023mud} do not contain the Fermi coupling constant and CKM matrix elements, so we cannot compare directly the respective topological amplitudes with theirs  except for the relative ratio. It is obvious that the relative sizes of the two topological amplitudes have different pattern in the two invariant amplitudes $H_S$ and $H_P$. The PQCD prediction on $|E/C|$ for $H_{S(P)}$ is smaller (larger) than that of SU(3). The calculated relative phases are around 2.3 rad, implies the interferences between $C$ and $E$ amplitudes are indeed destructive, and support the assumption made in Ref~\cite{Hsiao:2023mud}. Nevertheless, the somewhat deviation from $\pi$ in our calculations implies the destructive interferences are not at the maximum.

As stressed in~\cite{Hsiao:2023mud},  if the $W$-exchange contribution is dropped, the measured upper limit of $\mathcal{B}(\bar {B}^0\rightarrow\Lambda_c^+\bar {\Lambda}_c^-)$ is significantly below a naive extrapolation from $\mathcal{B}(B^-\rightarrow\bar {\Lambda}_c^-\Xi_c^0)$ assuming a simple Cabibbo-suppression factor of $|V_{cd}/V_{cs}|^2$. Then  they argued that a destructive interfering effect between the $W$-exchange and $W$-emission amplitudes can reduce the overestimated branching ratio to accommodate the measurement. However, this explanation seem to be not sufficient, because their central value still saturates the experimental upper bound slightly as  exhibited in Table~\ref{tab:branching} in spite of considering a maximum destructive interference. In this case, a way out is to consider SU(3) breaking effect since the small energy release in the doubly-charmed baryonic $B$ decays is not enough to weaken the differences among the  SU(3) multiplets.  At the hadron level, SU(3) flavor breaking due to the strange and light quark differences will manifest in the baryon masses, decay constants, and baryonic LCDAs in our calculations. When the $W$-exchange contribution is turned off, we obtain the central value of branching ratio $ \mathcal{B}(\bar {B}^0\rightarrow\Lambda_c^+\bar \Lambda_c^-)=1.3\times10^{-5}$, which is still below the upper bound. When the $W$-exchange  contribution is turned on, the calculated value further reduces to $8.8\times10^{-6}$ owing to the aforementioned destructive interference. We thereby point out that the SU(3) symmetry breaking as well as the interference effect play an essential role in understanding the measurements of $\mathcal{B}(\bar {B}^0\rightarrow\Lambda_c^+\bar \Lambda_c^-)$ and $ \mathcal{B}({B}^-\rightarrow\Xi_c^0\bar \Lambda_c^-)$. The significant SU(3) breaking effect can also explain  why our value is lower than the SU(3) one by a factor 2.

\begin{table}[!htbh]
    \caption{Asymmetry parameters for the considered decays. The theoretical uncertainties are the same as for branching ratios in Table~\ref{tab:branching}.} 	\label{tab:asys}
 \begin{tabular}[t]{lccc}
 \hline\hline
Mode  &$\alpha$ & $\beta$ &$\gamma$ \\ \hline
$ {B}^-\rightarrow\Xi_c^0\bar \Lambda_c^-$ &$-0.01^{+0.10+0.12+0.05+0.01}_{-0.10-0.29-0.14-0.01}$&$-0.99^{+0.01+0.09+0.00+0.00}_{-0.00-0.01-0.00-0.00}$&$-0.07^{+0.07+0.38+0.04+0.07}_{-0.06-0.13-0.05-0.08}$\\
$ \bar{B}^0_s\rightarrow\Lambda_c^+\bar \Lambda_c^-$ &$-0.03^{+0.05+0.03+0.05+0.01}_{-0.04-0.04-0.03-0.00}$&$-0.57^{+0.02+0.02+0.00+0.05}_{-0.03-0.02-0.02-0.05}$&$-0.82^{+0.03+0.02+0.01+0.04}_{-0.01-0.01-0.00-0.03}$\\
$ \bar{B}^0\rightarrow\Lambda_c^+\bar \Lambda_c^-$ &$0.17^{+0.08+0.08+0.03+0.02}_{-0.08-0.05-0.18-0.01}$&$-0.97^{+0.04+0.06+0.02+0.02}_{-0.03-0.00-0.02-0.01}$&$-0.15^{+0.17+0.54+0.14+0.09}_{-0.14-0.16-0.11-0.11}$\\
		\hline\hline
	\end{tabular}
\end{table}	

We next turn to some interesting asymmetry parameters in the decays under consideration, defined by~\cite{Geng:2023nia}
\begin{eqnarray}\label{eq:asym}
\alpha=\frac{|H_+|^2-|H_-|^2}{|H_+|^2+|H_-|^2}, \quad \beta=\frac{2Im(H_+H_-^*)}{|H_+|^2+|H_-|^2}, \quad  \gamma=\frac{2Re(H_+H_-^*)}{|H_+|^2+|H_-|^2},
\end{eqnarray}
with $H_{\pm}=\frac{1}{\sqrt{2}}(\sqrt{Q_+}H_S\mp\sqrt{Q_-}H_P)$. $\alpha$ is the up-down asymmetry parameter. $\beta$ is a naively $T$-odd observable. They are both $P$-odd, while $\gamma$ is $P$-even. They are related by  the identity $\alpha^2+ \beta^2+ \gamma^2=1$. Our results on these asymmetries are summarized in Table~\ref{tab:asys}, where the uncertainties are from the same quantities as above. The most important source of the theoretical errors is the assumption of the heavy quark limit for the heavy baryon LCDAs. We remark that the uncertainties due to the baryon LCDAs may be exaggerated because the parameters used here are actually for bottom baryons rather than charmed baryons, which are not known at present.  Any significant reduction of the error requires more accurate information on the charmed baryon LCDAs. As these asymmetries received less theoretical and experimental attentions, our predictions can be compared in future.

\section{ conclusion}\label{sec:sum}
Baryonic $B$ decays offer alternative robust ways to test the SM and search for new physics, complementing searches with  mesonic $B$-decays. However, the QCD dynamics of baryonic $B$ decay processes are more complicated than those of mesonic ones and poorly understood theoretically. In particular, the quantitative analyses on the baryonic $B$ decays based on the QCD-inspired approaches are still not developed.  In this work, we limit out attention to the four observed two-body doubly charmed baryonic $B$ decays and make the first steps to calculate their decay branching ratios and asymmetry parameters in the framework of PQCD. Our results could be helpful for a future experimental search for them and provide deeper insights in the understanding  of the dynamics of baryonic $B$ decay processes.

From the previous argument that the energy release in the doubled-charmed two-body decay is small, in order to improve the theoretical calculation accuracy, some higher-power corrections arise from the nonperturbative hadron distribution amplitudes are  taken into account in our numerical analysis. The subleading contribution of the $B$ meson distribution amplitude and the charmed baryon light-cone distribution amplitudes up to the twist-4 level are considered for the processes in question. It is found that the effect of the subleading component of the $B$ meson distribution amplitude is indeed essential and cannot be neglected. The subleading contributions can interfere destructively or constructively with the leading ones.  Contributions to the decay amplitude from various twist combinations of the baryon and antibaryon LCDAs are elaborated. It is clear that the higher-twist baryon LCDAs give the dominant contribution to the decay amplitudes of all the four modes.

With the decay amplitudes, we predict the decay branching ratios and asymmetry parameters, where the theoretical uncertainties arise from the nonperturbative hadronic LCDAs, the Sudakov resummation, and the hard scales. The accuracy of the theoretical predictions can be systematically improved once the charmed baryon LCDAs  are available in the future. Our prediction on the branching ratio of ${B}^-\rightarrow\Xi_c^0\bar \Lambda_c^-$  is compatible with the world average value,  while the values of $ \mathcal{B}(\bar {B}^0\rightarrow\Lambda_c^+\bar \Lambda_c^-)$ and $ \mathcal{B}(\bar {B}^0_s\rightarrow\Lambda_c^+\bar \Lambda_c^-)$  satisfy the existing experimental bounds. We also compare the calculated decay branching ratios with those from other theoretical approaches when they are available.  It is discovered that various results span wide ranges, which call for more precise measurements to discriminate them. The asymmetry parameters are investigated for the first time in this work. As the asymmetries have received less attentions in the literature, we await future comparisons.

We emphasize that the SU(3) breaking effects are crucial in explaining the measured branching ratios of $\mathcal{B}(\bar {B}^0\rightarrow\Lambda_c^+\bar \Lambda_c^-)$ and $ \mathcal{B}({B}^-\rightarrow\Xi_c^0\bar \Lambda_c^-)$. An explicit study of the  SU(3) breaking effects to other double charmed baryonic $B$ decays in PQCD is in progress.

\begin{acknowledgments}
We would like to acknowledge Professor Hsiang-nan Li and  Professor Yu-Kuo Hsiao for helpful discussions. This work is supported by National Natural Science Foundation of China under Grants Nos. 12075086, 12375089 and 12435004. Z.R. is also supported in part by the Natural Science Foundation of Hebei Province under Grant No. A2021209002 and No. A2019209449.  Z.T.Z. is  supported by  the Natural Science Foundation of Shandong province under the Grant No. ZR2022MA035. Y.Li is also supported by  the Natural Science Foundation of Shandong province under the Grant No. ZR2022ZD26.
\end{acknowledgments}

\begin{appendix}
\section{FACTORIZATION FORMULAS}\label{sec:for}
In this appendix, we give the expressions of those quantities associated with specific diagram entering to the decay amplitude in Eq.~(\ref{eq:ab}). The expressions of $a^{\sigma}$, $\Omega_{R_{ij}}$, and $t_{A,B,C,D}$ are given in Tables~\ref{tab:wilson},~\ref{tab:bb}, and~\ref{tab:ttt}, respectively. The auxiliary functions $h_l$ in Table~\ref{tab:bb} can be found in~\cite{220209181}.

In the present work, the formulas of $H^{\sigma}_{R_{ij}}(x,x',y)$ are rather lengthy due to many higher twist contributions are included. Here,  we only show some details for diagrams Fig.~\ref{fig:C} (b6) and  Fig.~\ref{fig:E} (c7), which represent the main contributions in each topology. The others can be derived in a similar way.

For the $S$-wave expressions of Fig.~\ref{fig:C} (b6):
\begin{eqnarray}
H^{LL,S}_{C_{b6}}(x,x',y)&=&\frac{M^3 }{f^--f^+}(\bar{r} x_1 x_2 (2 \phi _B (\bar{\phi }_3^a+\bar{\phi }_3^s-2 \bar{\phi }_4) \phi _3^s+2 \bar{\phi }_4 \bar{\phi }_B \phi
   _3^s+\phi _2 \phi _B \bar{\phi }_2+\phi _2 (\bar{\phi }_3^a-\bar{\phi }_3^s) (2 \phi _B-\bar{\phi }_B))
   (f^+)^2\nonumber\\&&+(r (2 \bar{\phi }_2 \bar{\phi }_B \phi _3^s-\phi _B (-2 \phi _3^s \bar{\phi }_3^a+\phi _4 \bar{\phi }_3^a+2 \phi _3^s
   \bar{\phi }_3^s+\phi _4 \bar{\phi }_3^s+2 \phi _3^s \bar{\phi }_2+\phi _4 \bar{\phi }_4)+\phi _4 (\bar{\phi }_3^a+\bar{\phi }_3^s)
   \bar{\phi }_B) (x'_2+x'_3)\nonumber\\&&-\bar{r} \bar{\phi }_B (x_2 (2 (\bar{\phi }_3^a+\bar{\phi }_3^s) \phi _3^s+\phi _2
   \bar{\phi }_2)+(2 \bar{\phi }_2 \phi _3^s+\phi _4 (\bar{\phi }_3^a+\bar{\phi }_3^s)) (x'_2+x'_3))\nonumber\\&&+\bar{r}
   \phi _B (2 x_2 (\bar{\phi }_3^a+\bar{\phi }_3^s+\bar{\phi }_4) \phi _3^s+x_2 \phi _2 (-\bar{\phi }_3^a+\bar{\phi }_3^s+\bar{\phi
   }_2)\nonumber\\&&+(-2 \phi _3^s \bar{\phi }_3^a+\phi _4 \bar{\phi }_3^a+2 \phi _3^s \bar{\phi }_3^s+\phi _4 \bar{\phi }_3^s+2 \phi _3^s \bar{\phi }_2+\phi
   _4 \bar{\phi }_4) (x'_2+x'_3))) f^+\nonumber\\&&+\bar{r} (2 \bar{\phi }_2 \bar{\phi }_B \phi _3^s-\phi _B ((\phi _4-2 \phi
   _3^s) \bar{\phi }_3^a+2 \phi _3^s \bar{\phi }_3^s+\phi _4 \bar{\phi }_3^s+2 \phi _3^s \bar{\phi }_2+\phi _4 \bar{\phi }_4)+\phi _4
   (\bar{\phi }_3^a+\bar{\phi }_3^s) \bar{\phi }_B) (x'_2+x'_3)\nonumber\\&&+f^- ((r-\bar{r}) x_1 x_2 (2 \phi _B
   (\bar{\phi }_3^a+\bar{\phi }_3^s-2 \bar{\phi }_4) \phi _3^s+2 \bar{\phi }_4 \bar{\phi }_B \phi _3^s+\phi _2 \phi _B \bar{\phi }_2+\phi _2
   (\bar{\phi }_3^a-\bar{\phi }_3^s) (2 \phi _B-\bar{\phi }_B)) (f^+)^2\nonumber\\&&+(r-\bar{r}) (x_2 \phi _B
   (2 (\bar{\phi }_3^a+\bar{\phi }_3^s+\bar{\phi }_4) \phi _3^s+\phi _2 (-\bar{\phi }_3^a+\bar{\phi }_3^s+\bar{\phi
   }_2))-x_2 (2 (\bar{\phi }_3^a+\bar{\phi }_3^s) \phi _3^s+\phi _2 \bar{\phi }_2) \bar{\phi }_B\nonumber\\&&+(2 \bar{\phi }_2
   (\phi _B-\bar{\phi }_B) \phi _3^s+\phi _B ((\phi _4-2 \phi _3^s) \bar{\phi }_3^a+2 \phi _3^s \bar{\phi }_3^s+\phi _4
   (\bar{\phi }_3^s+\bar{\phi }_4))-\phi _4 (\bar{\phi }_3^a+\bar{\phi }_3^s) \bar{\phi }_B) (x'_2+x'_3))
   f^+\nonumber\\&&+\bar{r} (2 \bar{\phi }_2 (\phi _B-\bar{\phi }_B) \phi _3^s+\phi _B ((\phi _4-2 \phi _3^s) \bar{\phi }_3^a+2 \phi _3^s
   \bar{\phi }_3^s+\phi _4 (\bar{\phi }_3^s+\bar{\phi }_4))-\phi _4 (\bar{\phi }_3^a+\bar{\phi }_3^s) \bar{\phi }_B)
   (x'_2+x'_3))),\nonumber\\&&
\end{eqnarray}
where $\bar \phi_{2,3,4}$  denote the corresponding quantities for an antibaryon.

\begin{eqnarray}
H^{SP,S}_{C_{b6}}(x,x',y)&=&-\frac{M^3}{2 (f^--f^+)} (\bar{r} x_1 x_2 (2 (\phi _B (\bar{\phi }_3^a+\bar{\phi }_3^s-2 \bar{\phi }_4)+\bar{\phi }_4 \bar{\phi }_B) \phi
   _3^s+\phi _2 \phi _B \bar{\phi }_2+\phi _2 (\bar{\phi }_3^a-\bar{\phi }_3^s) (2 \phi _B-\bar{\phi }_B)) (f^+)^2\nonumber\\&&+r
   ((2 \bar{\phi }_2 \phi _3^s+\phi _4 (\bar{\phi }_3^a+\bar{\phi }_3^s)) \bar{\phi }_B-\phi _B ((\phi _4-2 \phi
   _3^s) \bar{\phi }_3^a+2 \phi _3^s \bar{\phi }_3^s+\phi _4 \bar{\phi }_3^s+2 \phi _3^s \bar{\phi }_2+\phi _4 \bar{\phi }_4))
   (x'_2+x'_3) f^+\nonumber\\&&+\bar{r} (x_2 (2 \phi _B (\bar{\phi }_3^a+\bar{\phi }_3^s+\bar{\phi }_4) \phi _3^s-2 (\bar{\phi
   }_3^a+\bar{\phi }_3^s) \bar{\phi }_B \phi _3^s+\phi _2 \phi _B (-\bar{\phi }_3^a+\bar{\phi }_3^s+\bar{\phi }_2)-\phi _2 \bar{\phi }_2
   \bar{\phi }_B)\nonumber\\&&+(2 \bar{\phi }_2 (\phi _B-\bar{\phi }_B) \phi _3^s+\phi _B ((\phi _4-2 \phi _3^s) \bar{\phi }_3^a+2
   \phi _3^s \bar{\phi }_3^s+\phi _4 (\bar{\phi }_3^s+\bar{\phi }_4))-\phi _4 (\bar{\phi }_3^a+\bar{\phi }_3^s) \bar{\phi
   }_B) (x'_2+x'_3)) f^+\nonumber\\&&+\bar{r} ((2 \bar{\phi }_2 \phi _3^s+\phi _4 (\bar{\phi }_3^a+\bar{\phi }_3^s))
   \bar{\phi }_B-\phi _B ((\phi _4-2 \phi _3^s) \bar{\phi }_3^a+2 \phi _3^s \bar{\phi }_3^s+\phi _4 \bar{\phi }_3^s+2 \phi _3^s \bar{\phi
   }_2+\phi _4 \bar{\phi }_4)) (x'_2+x'_3)\nonumber\\&&+f^- ((r-\bar{r}) x_1 x_2 (2 (\phi _B (\bar{\phi
   }_3^a+\bar{\phi }_3^s-2 \bar{\phi }_4)+\bar{\phi }_4 \bar{\phi }_B) \phi _3^s+\phi _2 \phi _B \bar{\phi }_2+\phi _2 (\bar{\phi
   }_3^a-\bar{\phi }_3^s) (2 \phi _B-\bar{\phi }_B)) (f^+)^2\nonumber\\&&+(r-\bar{r}) (x_2 (2 \phi _B
   (\bar{\phi }_3^a+\bar{\phi }_3^s+\bar{\phi }_4) \phi _3^s-2 (\bar{\phi }_3^a+\bar{\phi }_3^s) \bar{\phi }_B \phi _3^s+\phi _2 \phi
   _B (-\bar{\phi }_3^a+\bar{\phi }_3^s+\bar{\phi }_2)-\phi _2 \bar{\phi }_2 \bar{\phi }_B)\nonumber\\&&+(2 \bar{\phi }_2 (\phi _B-\bar{\phi
   }_B) \phi _3^s+\phi _B ((\phi _4-2 \phi _3^s) \bar{\phi }_3^a+2 \phi _3^s \bar{\phi }_3^s+\phi _4 (\bar{\phi }_3^s+\bar{\phi
   }_4))-\phi _4 (\bar{\phi }_3^a+\bar{\phi }_3^s) \bar{\phi }_B) (x'_2+x'_3)) f^+\nonumber\\&&+\bar{r} (2 \bar{\phi
   }_2 (\phi _B-\bar{\phi }_B) \phi _3^s+\phi _B ((\phi _4-2 \phi _3^s) \bar{\phi }_3^a+2 \phi _3^s \bar{\phi }_3^s+\phi _4
   (\bar{\phi }_3^s+\bar{\phi }_4))-\phi _4 (\bar{\phi }_3^a+\bar{\phi }_3^s) \bar{\phi }_B)
   (x'_2+x'_3))). \nonumber\\&&
\end{eqnarray}

For the $S$-wave expressions of Fig.~\ref{fig:E} (c7):
\begin{eqnarray}
H^{LL,S}_{E_{c7}}(x,x',y)&=&\frac{M^3}{f^--f^+} ((\bar{r}+(r-\bar{r}) f^+) (\phi _B (2 \phi _3^a \bar{\phi }_3^a+2 \phi _3^s \bar{\phi }_3^s+\phi _2
   \bar{\phi }_2+2 \phi _4 \bar{\phi }_2+\phi _4 \bar{\phi }_4) \nonumber\\&&-(2 \phi _3^a \bar{\phi }_3^a+2 \phi _3^s \bar{\phi }_3^s+\phi _2 \bar{\phi
   }_2+\phi _4 \bar{\phi }_4) \bar{\phi }_B) x'_2 x'_3 f^{-2}+(-(r-\bar{r}) x_2 x_3 (\phi _B (2 \phi _3^a \bar{\phi
   }_3^a+2 \phi _3^s \bar{\phi }_3^s\nonumber\\&& +\phi _4 \bar{\phi }_4+\phi _2 (\bar{\phi }_2+2 \bar{\phi }_4))-2 \phi _2 \bar{\phi }_4 \bar{\phi
   }_B) f^{+2}-(r-\bar{r}) (-2 x_2 ((\phi _3^a+\phi _3^s) (\bar{\phi }_3^a-\bar{\phi }_3^s)+\phi _2
   \bar{\phi }_4) (\phi _B-\bar{\phi }_B)\nonumber\\&&+y ((2 \phi _3^a \bar{\phi }_3^a+2 \phi _3^s \bar{\phi }_3^s+\phi _2 \bar{\phi }_2+\phi
   _4 \bar{\phi }_4) \bar{\phi }_B-\phi _B (2 \phi _3^a \bar{\phi }_3^a+2 \phi _3^s \bar{\phi }_3^s+\phi _2 \bar{\phi }_2+2 \phi _4 \bar{\phi
   }_2+\phi _4 \bar{\phi }_4)) x'_3\nonumber\\&&-(2 \phi _3^a \bar{\phi }_3^a+2 \phi _3^s \bar{\phi }_3^s+\phi _2 \bar{\phi }_2+\phi _4 \bar{\phi
   }_4) \bar{\phi }_B x'_2 (y+2 x'_3)+\phi _B x'_2 (-2 \phi _4 \bar{\phi }_2-2 (\phi _3^a-\phi _3^s) (\bar{\phi
   }_3^a+\bar{\phi }_3^s)\nonumber\\&&+y (2 \phi _3^a \bar{\phi }_3^a+2 \phi _3^s \bar{\phi }_3^s+\phi _2 \bar{\phi }_2+2 \phi _4 \bar{\phi }_2+\phi _4
   \bar{\phi }_4)+2 (2 \phi _3^a \bar{\phi }_3^a+2 \phi _3^s \bar{\phi }_3^s+\phi _2 \bar{\phi }_2+2 \phi _4 \bar{\phi }_2+\phi _4 \bar{\phi
   }_4) x'_3)) f^+\nonumber\\&&-\bar{r} (y ((2 \phi _3^a \bar{\phi }_3^a+2 \phi _3^s \bar{\phi }_3^s+\phi _2 \bar{\phi }_2+\phi _4
   \bar{\phi }_4) \bar{\phi }_B-\phi _B (2 \phi _3^a \bar{\phi }_3^a+2 \phi _3^s \bar{\phi }_3^s+\phi _2 \bar{\phi }_2+2 \phi _4 \bar{\phi
   }_2+\phi _4 \bar{\phi }_4)) x'_3\nonumber\\&&-(2 \phi _3^a \bar{\phi }_3^a+2 \phi _3^s \bar{\phi }_3^s+\phi _2 \bar{\phi }_2+\phi _4 \bar{\phi
   }_4) \bar{\phi }_B x'_2 (y+2 x'_3)+\phi _B x'_2 (-2 \phi _4 \bar{\phi }_2-2 (\phi _3^a-\phi _3^s) (\bar{\phi
   }_3^a+\bar{\phi }_3^s)\nonumber\\&&+y (2 \phi _3^a \bar{\phi }_3^a+2 \phi _3^s \bar{\phi }_3^s+\phi _2 \bar{\phi }_2+2 \phi _4 \bar{\phi }_2+\phi _4
   \bar{\phi }_4)+2 (2 \phi _3^a \bar{\phi }_3^a+2 \phi _3^s \bar{\phi }_3^s+\phi _2 \bar{\phi }_2+2 \phi _4 \bar{\phi }_2+\phi _4 \bar{\phi
   }_4) x'_3))) f^-\nonumber\\&&-\bar{r} f^{+2} x_2 x_3 (\phi _B (2 \phi _3^a \bar{\phi }_3^a+2 \phi _3^s \bar{\phi }_3^s+\phi _4
   \bar{\phi }_4+\phi _2 (\bar{\phi }_2+2 \bar{\phi }_4))-2 \phi _2 \bar{\phi }_4 \bar{\phi }_B)\nonumber\\&&+\bar{r} (x'_2-y)
   (\phi _B (-2 \phi _4 \bar{\phi }_2-2 (\phi _3^a-\phi _3^s) (\bar{\phi }_3^a+\bar{\phi }_3^s)+y (2 \phi _3^a
   \bar{\phi }_3^a+2 \phi _3^s \bar{\phi }_3^s+\phi _2 \bar{\phi }_2+2 \phi _4 \bar{\phi }_2+\phi _4 \bar{\phi }_4)\nonumber\\&&+(2 \phi _3^a \bar{\phi
   }_3^a+2 \phi _3^s \bar{\phi }_3^s+\phi _2 \bar{\phi }_2+2 \phi _4 \bar{\phi }_2+\phi _4 \bar{\phi }_4) x'_3)-(2 \phi _3^a \bar{\phi
   }_3^a+2 \phi _3^s \bar{\phi }_3^s+\phi _2 \bar{\phi }_2+\phi _4 \bar{\phi }_4) \bar{\phi }_B (y+x'_3))\nonumber\\&&+f^+ (r
   (x'_2-y) (\phi _B (-2 \phi _4 \bar{\phi }_2-2 (\phi _3^a-\phi _3^s) (\bar{\phi }_3^a+\bar{\phi }_3^s)+y
   (2 \phi _3^a \bar{\phi }_3^a+2 \phi _3^s \bar{\phi }_3^s+\phi _2 \bar{\phi }_2+2 \phi _4 \bar{\phi }_2+\phi _4 \bar{\phi }_4)\nonumber\\&&+(2 \phi
   _3^a \bar{\phi }_3^a+2 \phi _3^s \bar{\phi }_3^s+\phi _2 \bar{\phi }_2+2 \phi _4 \bar{\phi }_2+\phi _4 \bar{\phi }_4) x'_3)-(2 \phi _3^a
   \bar{\phi }_3^a+2 \phi _3^s \bar{\phi }_3^s+\phi _2 \bar{\phi }_2+\phi _4 \bar{\phi }_4) \bar{\phi }_B (y+x'_3))\nonumber\\&&+\bar{r} (2
   x_2 ((\phi _3^a+\phi _3^s) (\bar{\phi }_3^a-\bar{\phi }_3^s)+\phi _2 \bar{\phi }_4) (\phi _B-\bar{\phi }_B)-y
   (2 \phi _3^a \bar{\phi }_3^a+2 \phi _3^s \bar{\phi }_3^s+\phi _2 \bar{\phi }_2+\phi _4 \bar{\phi }_4) \bar{\phi }_B
   (y+x'_3)\nonumber\\&&+(2 \phi _3^a \bar{\phi }_3^a+2 \phi _3^s \bar{\phi }_3^s+\phi _2 \bar{\phi }_2+\phi _4 \bar{\phi }_4) \bar{\phi }_B x'_2
   (y+x'_3)+y \phi _B (-2 \phi _4 \bar{\phi }_2-2 (\phi _3^a-\phi _3^s) (\bar{\phi }_3^a+\bar{\phi }_3^s)\nonumber\\&&+y (2
   \phi _3^a \bar{\phi }_3^a+2 \phi _3^s \bar{\phi }_3^s+\phi _2 \bar{\phi }_2+2 \phi _4 \bar{\phi }_2+\phi _4 \bar{\phi }_4)+(2 \phi _3^a
   \bar{\phi }_3^a+2 \phi _3^s \bar{\phi }_3^s+\phi _2 \bar{\phi }_2+2 \phi _4 \bar{\phi }_2+\phi _4 \bar{\phi }_4) x'_3)\nonumber\\&&-\phi _B x'_2 (-2
   \phi _4 \bar{\phi }_2-2 (\phi _3^a-\phi _3^s) (\bar{\phi }_3^a+\bar{\phi }_3^s)+y (2 \phi _3^a \bar{\phi }_3^a+2 \phi _3^s
   \bar{\phi }_3^s+\phi _2 \bar{\phi }_2+2 \phi _4 \bar{\phi }_2+\phi _4 \bar{\phi }_4)\nonumber\\&&+(2 \phi _3^a \bar{\phi }_3^a+2 \phi _3^s \bar{\phi
   }_3^s+\phi _2 \bar{\phi }_2+2 \phi _4 \bar{\phi }_2+\phi _4 \bar{\phi }_4) x'_3)))),
\end{eqnarray}
\begin{eqnarray}
H^{SP,S}_{E_{c7}}(x,x',y)&=&-\frac{M^3}{2 (f^--f^+)} ((\phi _B (2 \phi _3^a \bar{\phi }_3^a+2 \phi _3^s \bar{\phi }_3^s+\phi _2 \bar{\phi }_2+2 \phi _4 \bar{\phi }_2+\phi _4 \bar{\phi
   }_4)-(2 \phi _3^a \bar{\phi }_3^a+2 \phi _3^s \bar{\phi }_3^s+\phi _2 \bar{\phi }_2+\phi _4 \bar{\phi }_4) \nonumber\\&&\bar{\phi }_B) x'_2
   x'_3 (\bar{r}+(r-\bar{r}) f^+) (f^-)^2+(-(r-\bar{r}) x_2 x_3 (\phi _B (2 \phi _3^a \bar{\phi
   }_3^a+2 \phi _3^s \bar{\phi }_3^s+\phi _4 \bar{\phi }_4+\phi _2 (\bar{\phi }_2+2 \bar{\phi }_4))\nonumber\\&&-2 \phi _2 \bar{\phi }_4 \bar{\phi
   }_B) (f^+)^2-(r-\bar{r}) (-2 x_2 ((\phi _3^a+\phi _3^s) (\bar{\phi }_3^a-\bar{\phi
   }_3^s)+\phi _2 \bar{\phi }_4) (\phi _B-\bar{\phi }_B)+y ((2 \phi _3^a \bar{\phi }_3^a+2 \phi _3^s \bar{\phi }_3^s\nonumber\\&&+\phi
   _2 \bar{\phi }_2+\phi _4 \bar{\phi }_4) \bar{\phi }_B-\phi _B (2 \phi _3^a \bar{\phi }_3^a+2 \phi _3^s \bar{\phi }_3^s+\phi _2 \bar{\phi }_2+2
   \phi _4 \bar{\phi }_2+\phi _4 \bar{\phi }_4)) x'_3-(2 \phi _3^a \bar{\phi }_3^a+2 \phi _3^s \bar{\phi }_3^s\nonumber\\&&+\phi _2 \bar{\phi }_2+\phi
   _4 \bar{\phi }_4) \bar{\phi }_B x'_2 (y+2 x'_3)+\phi _B x'_2 (-2 \phi _4 \bar{\phi }_2-2 (\phi _3^a-\phi _3^s)
   (\bar{\phi }_3^a+\bar{\phi }_3^s)+y (2 \phi _3^a \bar{\phi }_3^a+2 \phi _3^s \bar{\phi }_3^s\nonumber\\&&+\phi _2 \bar{\phi }_2+2 \phi _4 \bar{\phi
   }_2+\phi _4 \bar{\phi }_4)+2 (2 \phi _3^a \bar{\phi }_3^a+2 \phi _3^s \bar{\phi }_3^s+\phi _2 \bar{\phi }_2+2 \phi _4 \bar{\phi }_2+\phi _4
   \bar{\phi }_4) x'_3)) f^+\nonumber\\&&-\bar{r} (y ((2 \phi _3^a \bar{\phi }_3^a+2 \phi _3^s \bar{\phi }_3^s+\phi _2 \bar{\phi
   }_2+\phi _4 \bar{\phi }_4) \bar{\phi }_B-\phi _B (2 \phi _3^a \bar{\phi }_3^a+2 \phi _3^s \bar{\phi }_3^s+\phi _2 \bar{\phi }_2+2 \phi _4
   \bar{\phi }_2+\phi _4 \bar{\phi }_4)) x'_3\nonumber\\&&-(2 \phi _3^a \bar{\phi }_3^a+2 \phi _3^s \bar{\phi }_3^s+\phi _2 \bar{\phi }_2+\phi _4
   \bar{\phi }_4) \bar{\phi }_B x'_2 (y+2 x'_3)+\phi _B x'_2 (-2 \phi _4 \bar{\phi }_2-2 (\phi _3^a-\phi _3^s)
   (\bar{\phi }_3^a+\bar{\phi }_3^s)\nonumber\\&&+y (2 \phi _3^a \bar{\phi }_3^a+2 \phi _3^s \bar{\phi }_3^s+\phi _2 \bar{\phi }_2+2 \phi _4 \bar{\phi
   }_2+\phi _4 \bar{\phi }_4)+2 (2 \phi _3^a \bar{\phi }_3^a+2 \phi _3^s \bar{\phi }_3^s+\phi _2 \bar{\phi }_2+2 \phi _4 \bar{\phi }_2+\phi _4
   \bar{\phi }_4) x'_3))) f^-\nonumber\\&&-\bar{r} x_2 x_3 (\phi _B (2 \phi _3^a \bar{\phi }_3^a+2 \phi _3^s \bar{\phi }_3^s+\phi _4
   \bar{\phi }_4+\phi _2 (\bar{\phi }_2+2 \bar{\phi }_4))-2 \phi _2 \bar{\phi }_4 \bar{\phi }_B) (f^+)^2\nonumber\\&&+\bar{r}
   (x'_2-y) (\phi _B (-2 \phi _4 \bar{\phi }_2-2 (\phi _3^a-\phi _3^s) (\bar{\phi }_3^a+\bar{\phi }_3^s)+y
   (2 \phi _3^a \bar{\phi }_3^a+2 \phi _3^s \bar{\phi }_3^s+\phi _2 \bar{\phi }_2+2 \phi _4 \bar{\phi }_2+\phi _4 \bar{\phi }_4)\nonumber\\&&+(2 \phi
   _3^a \bar{\phi }_3^a+2 \phi _3^s \bar{\phi }_3^s+\phi _2 \bar{\phi }_2+2 \phi _4 \bar{\phi }_2+\phi _4 \bar{\phi }_4) x'_3)-(2 \phi _3^a
   \bar{\phi }_3^a+2 \phi _3^s \bar{\phi }_3^s+\phi _2 \bar{\phi }_2+\phi _4 \bar{\phi }_4) \bar{\phi }_B (y+x'_3))\nonumber\\&&+(r
   (x'_2-y) (\phi _B (-2 \phi _4 \bar{\phi }_2-2 (\phi _3^a-\phi _3^s) (\bar{\phi }_3^a+\bar{\phi }_3^s)+y
   (2 \phi _3^a \bar{\phi }_3^a+2 \phi _3^s \bar{\phi }_3^s+\phi _2 \bar{\phi }_2+2 \phi _4 \bar{\phi }_2+\phi _4 \bar{\phi }_4)\nonumber\\&&+(2 \phi
   _3^a \bar{\phi }_3^a+2 \phi _3^s \bar{\phi }_3^s+\phi _2 \bar{\phi }_2+2 \phi _4 \bar{\phi }_2+\phi _4 \bar{\phi }_4) x'_3)-(2 \phi _3^a
   \bar{\phi }_3^a+2 \phi _3^s \bar{\phi }_3^s+\phi _2 \bar{\phi }_2+\phi _4 \bar{\phi }_4) \bar{\phi }_B (y+x'_3))\nonumber\\&&+\bar{r} (2
   x_2 ((\phi _3^a+\phi _3^s) (\bar{\phi }_3^a-\bar{\phi }_3^s)+\phi _2 \bar{\phi }_4) (\phi _B-\bar{\phi }_B)-y
   (2 \phi _3^a \bar{\phi }_3^a+2 \phi _3^s \bar{\phi }_3^s+\phi _2 \bar{\phi }_2+\phi _4 \bar{\phi }_4) \bar{\phi }_B
   (y+x'_3)\nonumber\\&&+(2 \phi _3^a \bar{\phi }_3^a+2 \phi _3^s \bar{\phi }_3^s+\phi _2 \bar{\phi }_2+\phi _4 \bar{\phi }_4) \bar{\phi }_B x'_2
   (y+x'_3)+y \phi _B (-2 \phi _4 \bar{\phi }_2-2 (\phi _3^a-\phi _3^s) (\bar{\phi }_3^a+\bar{\phi }_3^s)\nonumber\\&&+y (2
   \phi _3^a \bar{\phi }_3^a+2 \phi _3^s \bar{\phi }_3^s+\phi _2 \bar{\phi }_2+2 \phi _4 \bar{\phi }_2+\phi _4 \bar{\phi }_4)+(2 \phi _3^a
   \bar{\phi }_3^a+2 \phi _3^s \bar{\phi }_3^s+\phi _2 \bar{\phi }_2+2 \phi _4 \bar{\phi }_2+\phi _4 \bar{\phi }_4) x'_3)\nonumber\\&&-\phi _B x'_2 (-2
   \phi _4 \bar{\phi }_2-2 (\phi _3^a-\phi _3^s) (\bar{\phi }_3^a+\bar{\phi }_3^s)+y (2 \phi _3^a \bar{\phi }_3^a+2 \phi _3^s
   \bar{\phi }_3^s+\phi _2 \bar{\phi }_2+2 \phi _4 \bar{\phi }_2+\phi _4 \bar{\phi }_4)\nonumber\\&&+(2 \phi _3^a \bar{\phi }_3^a+2 \phi _3^s \bar{\phi
   }_3^s+\phi _2 \bar{\phi }_2+2 \phi _4 \bar{\phi }_2+\phi _4 \bar{\phi }_4) x'_3))) f^+).
\end{eqnarray}

The corresponding formulas for the $P$-wave ones can be obtained by the following replacement:
\begin{eqnarray}
H^{LL,P}=H^{LL,S}|_{r\rightarrow -r}, \quad H^{SP,P}=H^{SP,S}|_{\bar r\rightarrow -\bar r}.
\end{eqnarray}

\begin{table} [H]
\footnotesize
	\caption{The expressions of $a^{LL}$ and $a^{SP}$  for the $W$-emission and $W$-exchange diagrams.} 
	\newcommand{\tabincell}[2]{\begin{tabular}{@{}#1@{}}#2\end{tabular}}
	\label{tab:wilson}
	\begin{tabular}[t]{lcc}
		\hline\hline
		$R_{ij}$        &$a^{LL}$       &$a^{SP}$             \\ \hline
		$C_{a1,a2,a4,a5,b1,b4,c1,c3,d1,d3,e1,f1}$        &$V_{cb}V^*_{cq}(C_2-C_1)-V_{tb}V^*_{tq}(C_4-C_3+C_{10}-C_9)$       &$-V_{tb}V^*_{tq}(C_6-C_5+C_{8}-C_7)$            \\
		
		$C_{a3,b2,c2,f2}$        &$V_{cb}V^*_{cq}(2C_1+C_2)-V_{tb}V^*_{tq}(2C_3+C_4+2C_9+C_{10})$       &$-V_{tb}V^*_{tq}(2C_5+C_6+2C_7+C_8)$            \\
		
		$C_{a6,b6,c5,d5,f3}$     &$V_{cb}V^*_{cq}2(C_1-C_2)-V_{tb}V^*_{tq}2(C_3-C_4+C_9-C_{10})$       &$-V_{tb}V^*_{tq}2(C_5-C_6+C_7-C_8)$            \\
		
		$C_{a7,b7,c6,f4}$        &$V_{cb}V^*_{cq}(-C_1-2C_2)-V_{tb}V^*_{tq}(-C_3-2C_4-C_9-2C_{10})$       &$-V_{tb}V^*_{tq}(-C_5-2C_6-C_7-2C_8)$            \\
		
		$C_{b3,d2,e2}$        &$V_{cb}V^*_{cq}(-\frac{1}{4}C_1+C_2)-V_{tb}V^*_{tq}(-\frac{1}{4}C_3+C_4-\frac{1}{4}C_9+C_{10})$       &$-V_{tb}V^*_{tq}(-\frac{1}{4}C_5+C_6-\frac{1}{4}C_7+C_8)$            \\
		
		$C_{b5}$        &$V_{cb}V^*_{cq}(\frac{5}{4}C_1+C_2)-V_{tb}V^*_{tq}(\frac{5}{4}C_3+C_4+\frac{5}{4}C_9+C_{10})$       &$-V_{tb}V^*_{tq}(\frac{5}{4}C_5+C_6+\frac{5}{4}C_7+C_8)$            \\
		
		$C_{c4}$        &$V_{cb}V^*_{cq}(-C_1-\frac{5}{4}C_2)-V_{tb}V^*_{tq}(-C_3-\frac{5}{4}C_4-C_9-\frac{5}{4}C_{10})$           &$-V_{tb}V^*_{tq}(-C_5-\frac{5}{4}C_6-C_7-\frac{5}{4}C_8)$            \\
		
		$C_{c7,d7,e4}$        &$V_{cb}V^*_{cq}(\frac{1}{4}C_2-C_1)-V_{tb}V^*_{tq}(\frac{1}{4}C_4-C_3+\frac{1}{4}C_{10}-C_9)$       &$-V_{tb}V^*_{tq}(\frac{1}{4}C_6-C_5+\frac{1}{4}C_8-C_7)$            \\
		
		$C_{d4}$        &$V_{cb}V^*_{cq}\frac{5}{4}(C_1-C_2)-V_{tb}V^*_{tq}\frac{5}{4}(C_3-C_4+C_9-C_{10})$       &$-V_{tb}V^*_{tq}\frac{5}{4}(C_5-C_6+C_7-C_8)$            \\
		
		$C_{d6}$     &$V_{cb}V^*_{cq}4(C_2-C_1)-V_{tb}V^*_{tq}4(C_4-C_3+C_{10}-C_9)$       &$-V_{tb}V^*_{tq}4(C_6-C_5+C_8-C_7)$            \\

        $C_{e3}$     &$V_{cb}V^*_{cq}(C_2-C_1)/4-V_{tb}V^*_{tq}(C_4-C_3+C_{10}-C_9)/4$       &$-V_{tb}V^*_{tq}(C_6-C_5+C_8-C_7)/4$            \\
		
		$C_{g1}$        &0       &0           \\

        $C_{g2}$        &$V_{cb}V^*_{cq}\frac{9}{4}(-C_1)-V_{tb}V^*_{tq}\frac{9}{4}(-C_3-C_9)$       &$-V_{tb}V^*_{tq}\frac{9}{4}(-C_5-C_7)$            \\
		
		$C_{g3}$        &$V_{cb}V^*_{cq}\frac{9}{4}(C_2-C_1)-V_{tb}V^*_{tq}\frac{9}{4}(C_4-C_3+C_{10}-C_9)$       &$-V_{tb}V^*_{tq}\frac{9}{4}(C_6-C_5+C_8-C_7)$            \\
		
		$C_{g4}$        &$V_{cb}V^*_{cq}\frac{9}{4}C_2-V_{tb}V^*_{tq}\frac{9}{4}(C_4+C_{10})$       &$-V_{tb}V^*_{tq}\frac{9}{4}(C_6+C_8)$            \\
		
		$E_{a1,a2,a3,a5,b1,b2,b4}$        &$V_{cb}V^*_{cq}(3C_1+C_2)-V_{tb}V^*_{tq}(3C_3+C_4+3C_9+C_{10})$       &$-V_{tb}V^*_{tq}(3C_5+C_6+3C_7+C_8)$            \\
		
		$E_{a6,a7,b6,b7,c1,c2,d1,d2}$        &$V_{cb}V^*_{cq}C_2-V_{tb}V^*_{tq}(C_4+C_{10})$       &$-V_{tb}V^*_{tq}(C_6+C_8)$            \\
		
		$E_{c5,c7,d6}$        &$V_{cb}V^*_{cq}(C_2+\frac{3}{4}C_1)-V_{tb}V^*_{tq}(C_4-\frac{3}{4}C_3+C_{10}-\frac{3}{4}C_9)$       &$-V_{tb}V^*_{tq}(C_6-\frac{3}{4}C_8+C_8-\frac{3}{4}C_7)$            \\	
		\hline\hline
	\end{tabular}
\end{table}

\begin{table}[H]
	\footnotesize
	\caption{The expressions of  $\Omega_{R_{ij}}$   for the $W$-emission and $W$-exchange diagrams.}
	\newcommand{\tabincell}[2]{\begin{tabular}{@{}#1@{}}#2\end{tabular}}
	\label{tab:bb}
	\begin{tabular}[t]{lcc}
		\hline\hline
$R_{ij}$  &$\Omega_{R_{ij}}$\\ \hline
		$C_{a1}$  &$K_0(\sqrt{t_A}|\textbf{b}_3|)h_2(|\textbf{b}_3+\textbf{b}'_2-\textbf{b}_2|,|\textbf{b}'_2|,t_C,t_B,t_D)\delta^2(\textbf{b}'_3)\delta^2(\textbf{b}_q-\textbf{b}_2+\textbf{b}'_2)$\\
		
		$C_{a2}$  &$K_0(\sqrt{t_B}|\textbf{b}_2-\textbf{b}_3|)h_2(|\textbf{b}'_2-\textbf{b}_2|,|\textbf{b}_3-\textbf{b}_2+\textbf{b}'_2|,t_A,t_C,t_D)\delta^2(\textbf{b}'_3)\delta^2(\textbf{b}_q-\textbf{b}_2+\textbf{b}'_2)$\\
		
		$C_{a3}$  &$\frac{1}{(2\pi)^3}K_0(\sqrt{t_A}|\textbf{b}'_2-\textbf{b}'_3-\textbf{b}_2|)K_0(\sqrt{t_B}|\textbf{b}_2-\textbf{b}_3|)K_0(\sqrt{t_C}|\textbf{b}'_3|)K_0(\sqrt{t_D}|\textbf{b}_3-\textbf{b}_2+\textbf{b}'_2|)\delta^2(\textbf{b}_q-\textbf{b}_2+\textbf{b}'_2)$\\
		
		$C_{a4}$  &$\frac{1}{(2\pi)^3}K_0(\sqrt{t_A}|\textbf{b}_2|)K_0(\sqrt{t_B}|\textbf{b}'_2-\textbf{b}_3+\textbf{b}_q|)K_0(\sqrt{t_C}|\textbf{b}_2-\textbf{b}'_2-\textbf{b}_q|)K_0(\sqrt{t_D}|\textbf{b}_3-\textbf{b}_q|)\delta^2(\textbf{b}'_3)$\\
		
		$C_{a5}$  &$\frac{1}{(2\pi)^3}K_0(\sqrt{t_A}|\textbf{b}'_2+\textbf{b}_q|)K_0(\sqrt{t_B}|\textbf{b}_2-\textbf{b}_3|)K_0(\sqrt{t_C}|\textbf{b}_2-\textbf{b}'_2-\textbf{b}_q|)K_0(\sqrt{t_D}|\textbf{b}_3-\textbf{b}_q|)\delta^2(\textbf{b}'_3)$\\
		
		$C_{a6}$  &$\frac{1}{(2\pi)^3}K_0(\sqrt{t_A}|\textbf{b}_q|)K_0(\sqrt{t_B}|\textbf{b}_2-\textbf{b}_3|)K_0(\sqrt{t_C}|\textbf{b}_2-\textbf{b}'_2-\textbf{b}_q|)K_0(\sqrt{t_D}|\textbf{b}_3-\textbf{b}_2+\textbf{b}'_2|)\delta^2(\textbf{b}'_3)$\\
		
		$C_{a7}$  &$\frac{1}{(2\pi)^3}K_0(\sqrt{t_A}|\textbf{b}'_2-\textbf{b}_2|)K_0(\sqrt{t_B}|\textbf{b}_2-\textbf{b}_3|)K_0(\sqrt{t_C}|\textbf{b}_2-\textbf{b}'_2-\textbf{b}_q|)K_0(\sqrt{t_D}|\textbf{b}_3-\textbf{b}_q|)\delta^2(\textbf{b}_2-\textbf{b}'_2+\textbf{b}'_3-\textbf{b}_q)$\\
		
		$C_{b1}$  &$\frac{1}{(2\pi)^3}K_0(\sqrt{t_A}|\textbf{b}_3|)K_0(\sqrt{t_B}|\textbf{b}'_2-\textbf{b}'_3|)K_0(\sqrt{t_C}|\textbf{b}_3-\textbf{b}_2+\textbf{b}'_2|)K_0(\sqrt{t_D}|\textbf{b}'_3|)\delta^2(\textbf{b}'_2-\textbf{b}_2+\textbf{b}_q)$\\
		
		$C_{b2}$  &$K_0(\sqrt{t_B}|\textbf{b}_2-\textbf{b}_3-\textbf{b}'_3|)h_2(|\textbf{b}_3|,|\textbf{b}'_3|,t_A,t_D,t_C)\delta^2(\textbf{b}_3-\textbf{b}_2+\textbf{b}'_2)\delta^2(\textbf{b}_q-\textbf{b}_3)$\\
		
		$C_{b3}$  &$K_0(\sqrt{t_A}|\textbf{b}_3+\textbf{b}'_3|) h_2(|\textbf{b}'_3|,|\textbf{b}_2-\textbf{b}_3|,t_C,t_B,t_D) \delta^2(\textbf{b}_3-\textbf{b}_2+\textbf{b}'_2) \delta^2(\textbf{b}_q-\textbf{b}_3)$\\
		
		$C_{b4}$  &$\frac{1}{(2\pi)^3}K_0(\sqrt{t_A}|\textbf{b}_2|) K_0(\sqrt{t_B}|\textbf{b}'_2-\textbf{b}'_3|) K_0(\sqrt{t_C}|\textbf{b}_2-\textbf{b}_3-\textbf{b}'_2|) K_0(\sqrt{t_D}|\textbf{b}'_3|) \delta^2(\textbf{b}_q-\textbf{b}_3)$\\
		
		$C_{b5}$  &$\frac{1}{(2\pi)^3}K_0(\sqrt{t_A}|\textbf{b}_3-\textbf{b}'_2|) K_0(\sqrt{t_B}|\textbf{b}_2-\textbf{b}_3-\textbf{b}'_3|) K_0(\sqrt{t_C}|\textbf{b}_3-\textbf{b}_2+\textbf{b}'_2|) K_0(\sqrt{t_D}|\textbf{b}'_3|) \delta^2(\textbf{b}_q-\textbf{b}_3)$\\
		
		$C_{b6}$  &$\frac{1}{(2\pi)^3}K_0(\sqrt{t_A}|\textbf{b}_q|) K_0(\sqrt{t_B}|\textbf{b}_2-\textbf{b}_3-\textbf{b}'_3|) K_0(\sqrt{t_C}|\textbf{b}_3-\textbf{b}_q|) K_0(\sqrt{t_D}|\textbf{b}'_3|) \delta^2(\textbf{b}_3-\textbf{b}_2+\textbf{b}'_2)$\\
		
		$C_{b7}$  &$\frac{1}{(2\pi)^3}K_0(\sqrt{t_A}|\textbf{b}'_2-\textbf{b}_2|) K_0(\sqrt{t_B}|\textbf{b}'_2-\textbf{b}'_3|) K_0(\sqrt{t_C}|\textbf{b}_2-\textbf{b}_3-\textbf{b}'_2|) K_0(\sqrt{t_D}|\textbf{b}_2-\textbf{b}_3-\textbf{b}'_2+\textbf{b}'_3|) \delta^2(\textbf{b}_q-\textbf{b}_3)$\\
		
		$C_{c1}$  &$\frac{1}{(2\pi)^3}K_0(\sqrt{t_A}|\textbf{b}_3|) K_0(\sqrt{t_B}|\textbf{b}'_2|) K_0(\sqrt{t_C}|\textbf{b}_3-\textbf{b}_2+\textbf{b}'_2-\textbf{b}'_3|) K_0(\sqrt{t_D}|\textbf{b}'_3|) \delta^2(\textbf{b}_q-\textbf{b}_2+\textbf{b}'_2-\textbf{b}'_3)$\\
		
		$C_{c2}$  &$\frac{1}{(2\pi)^3}K_0(\sqrt{t_A}|\textbf{b}'_2-\textbf{b}'_3-\textbf{b}_2|)  K_0(\sqrt{t_B}|\textbf{b}'_2|) K_0(\sqrt{t_C}|\textbf{b}_2-\textbf{b}_3-\textbf{b}'_2+\textbf{b}'_3|) K_0(\sqrt{t_D}|\textbf{b}_2-\textbf{b}_3-\textbf{b}'_2|) \delta^2(\textbf{b}_q-\textbf{b}_3)$\\
		
		$C_{c3}$  &$\frac{1}{(2\pi)^3}K_0(\sqrt{t_A}|\textbf{b}_2|)  K_0(\sqrt{t_B}|\textbf{b}'_2|) K_0(\sqrt{t_C}|\textbf{b}_2-\textbf{b}_3-\textbf{b}'_2+\textbf{b}'_3|) K_0(\sqrt{t_D}|\textbf{b}'_3|) \delta^2(\textbf{b}_q-\textbf{b}_3)$\\
		
		$C_{c4}$  &$\frac{1}{(2\pi)^3}K_0(\sqrt{t_A}|\textbf{b}'_3-\textbf{b}'_2-\textbf{b}_3|)  K_0(\sqrt{t_B}|\textbf{b}_2-\textbf{b}_3+\textbf{b}'_3|) K_0(\sqrt{t_C}|\textbf{b}_3-\textbf{b}_2+\textbf{b}'_2-\textbf{b}'_3|) K_0(\sqrt{t_D}|\textbf{b}'_3|) \delta^2(\textbf{b}_q-\textbf{b}_3)$\\
		
		$C_{c5}$  &$\frac{1}{(2\pi)^3}K_0(\sqrt{t_A}|\textbf{b}_q|)  K_0(\sqrt{t_B}|\textbf{b}'_2|) K_0(\sqrt{t_C}|\textbf{b}_3-\textbf{b}_q|) K_0(\sqrt{t_D}|\textbf{b}_2-\textbf{b}_3-\textbf{b}'_2|) \delta^2(\textbf{b}_2-\textbf{b}_3-\textbf{b}'_2+\textbf{b}'_3)$\\
		
		$C_{c6}$  &$K_0(\sqrt{t_B}|\textbf{b}'_2|) h_2(|\textbf{b}_3|,|\textbf{b}_2-\textbf{b}_3-\textbf{b}'_2|,t_A,t_D,t_C) \delta^2(\textbf{b}_2-\textbf{b}_3-\textbf{b}'_2+\textbf{b}'_3) \delta^2(\textbf{b}_q-\textbf{b}_3)$\\
		
		$C_{c7}$  &$K_0(\sqrt{t_A}|\textbf{b}'_2-\textbf{b}_2|) h_2(|\textbf{b}_2-\textbf{b}_3-\textbf{b}'_2|,|\textbf{b}_2-\textbf{b}_3|,t_C,t_B,t_D) \delta^2(\textbf{b}_2-\textbf{b}_3-\textbf{b}'_2+\textbf{b}'_3) \delta^2(\textbf{b}_q-\textbf{b}_3)$\\
		
		$C_{d1}$  &$\frac{1}{(2\pi)^3}K_0(\sqrt{t_A}|\textbf{b}_3|)  K_0(\sqrt{t_B}|\textbf{b}_2-\textbf{b}_q|) K_0(\sqrt{t_C}|\textbf{b}_3-\textbf{b}_2+\textbf{b}'_2|) K_0(\sqrt{t_D}|\textbf{b}_2-\textbf{b}'_2-\textbf{b}_q|) \delta^2(\textbf{b}'_3)$\\
		
		$C_{d2}$  &$\frac{1}{(2\pi)^3}K_0(\sqrt{t_A}|\textbf{b}_3+\textbf{b}'_3|)  K_0(\sqrt{t_B}|\textbf{b}_2-\textbf{b}_q|) K_0(\sqrt{t_C}|\textbf{b}'_3|) K_0(\sqrt{t_D}|\textbf{b}_3-\textbf{b}_q|) \delta^2(\textbf{b}_3-\textbf{b}_2+\textbf{b}'_2)$\\
		
		$C_{d3}$  &$\frac{1}{(2\pi)^3}K_0(\sqrt{t_A}|\textbf{b}_2|)  K_0(\sqrt{t_B}|\textbf{b}_3+\textbf{b}'_2-\textbf{b}_q|) K_0(\sqrt{t_C}|\textbf{b}_2-\textbf{b}_3-\textbf{b}'_2|) K_0(\sqrt{t_D}|\textbf{b}_3-\textbf{b}_q|) \delta^2(\textbf{b}'_3)$\\
		
		$C_{d4}$  &$\frac{1}{(2\pi)^3}K_0(\sqrt{t_A}|\textbf{b}_3+\textbf{b}'_2|)  K_0(\sqrt{t_B}|\textbf{b}_2-\textbf{b}_q|) K_0(\sqrt{t_C}|\textbf{b}_3-\textbf{b}_2+\textbf{b}'_2|) K_0(\sqrt{t_D}|\textbf{b}_3-\textbf{b}_q|) \delta^2(\textbf{b}'_3)$\\
		
		$C_{d5}$  &$K_0(\sqrt{t_A}|\textbf{b}_q|) h_2(|\textbf{b}_3-\textbf{b}_q|,|\textbf{b}_3-\textbf{b}_2|,t_C,t_B,t_D) \delta^2(\textbf{b}_3-\textbf{b}_2+\textbf{b}'_2) \delta^2(\textbf{b}'_3)$\\
		
		$C_{d6}$  &$K_0(\sqrt{t_B}|\textbf{b}_2-\textbf{b}_q|) h_2(|\textbf{b}_q|,|\textbf{b}_3-\textbf{b}_q|,t_A,t_C,t_D) \delta^2(\textbf{b}_3-\textbf{b}_2+\textbf{b}'_2) \delta^2(\textbf{b}'_3)$\\
		
		$C_{d7}$  &$\frac{1}{(2\pi)^3}K_0(\sqrt{t_A}|\textbf{b}'_2-\textbf{b}_2|)  K_0(\sqrt{t_B}|\textbf{b}_2-\textbf{b}_q|) K_0(\sqrt{t_C}|\textbf{b}_2-\textbf{b}_3-\textbf{b}'_2|) K_0(\sqrt{t_D}|\textbf{b}_3-\textbf{b}_q|) \delta^2(\textbf{b}_2-\textbf{b}_3-\textbf{b}'_2+\textbf{b}'_3)$\\
		
		$C_{e1}$  &$\frac{1}{(2\pi)^3}K_0(\sqrt{t_A}|\textbf{b}'_2+\textbf{b}_q|)  K_0(\sqrt{t_B}|\textbf{b}_3|) K_0(\sqrt{t_C}|\textbf{b}'_2-\textbf{b}_2+\textbf{b}_q|) K_0(\sqrt{t_D}|\textbf{b}_3-\textbf{b}_2+\textbf{b}'_2|) \delta^2(\textbf{b}'_3)$\\
		
		$C_{e2}$  &$\frac{1}{(2\pi)^3}K_0(\sqrt{t_A}|\textbf{b}_2-\textbf{b}_3+\textbf{b}_q|)  K_0(\sqrt{t_B}|\textbf{b}_3+\textbf{b}'_3|) K_0(\sqrt{t_C}|\textbf{b}_q-\textbf{b}_3|) K_0(\sqrt{t_D}|\textbf{b}'_3|) \delta^2(\textbf{b}_3-\textbf{b}_2+\textbf{b}'_2)$\\
		
		$C_{e3}$  &$h_3(|\textbf{b}_3|,|\textbf{b}_q|,|\textbf{b}_3+\textbf{b}'_2+\textbf{b}_q|,t_D,t_C,t_A,t_B) \delta^2(\textbf{b}_2-\textbf{b}_3-\textbf{b}'_2) \delta^2(\textbf{b}'_3)$\\
		
		$C_{e4}$  &$\frac{1}{(2\pi)^3}K_0(\sqrt{t_A}|\textbf{b}_2-\textbf{b}_3+\textbf{b}_q|)  K_0(\sqrt{t_B}|\textbf{b}_2-\textbf{b}'_2|) K_0(\sqrt{t_C}|\textbf{b}_q-\textbf{b}_3|) K_0(\sqrt{t_D}|\textbf{b}_2-\textbf{b}_3-\textbf{b}'_2|) \delta^2(\textbf{b}_2-\textbf{b}_3-\textbf{b}'_2+\textbf{b}'_3)$\\
		
		$C_{f1}$  &$\frac{1}{(2\pi)^3}K_0(\sqrt{t_A}|\textbf{b}_2|)  K_0(\sqrt{t_B}|\textbf{b}_3-\textbf{b}_2+\textbf{b}'_2+\textbf{b}_q|) K_0(\sqrt{t_C}|\textbf{b}'_2-\textbf{b}_2+\textbf{b}_q|) K_0(\sqrt{t_D}|\textbf{b}_3-\textbf{b}_2+\textbf{b}'_2|) \delta^2(\textbf{b}'_3)$\\
		
		$C_{f2}$  &$\frac{1}{(2\pi)^3}K_0(\sqrt{t_A}|\textbf{b}_2|)  K_0(\sqrt{t_B}|\textbf{b}'_3+\textbf{b}_q|) K_0(\sqrt{t_C}|\textbf{b}_q-\textbf{b}_3|) K_0(\sqrt{t_D}|\textbf{b}'_3|) \delta^2(\textbf{b}_3-\textbf{b}_2+\textbf{b}'_2)$\\
		
		$C_{f3}$  &$K_0(\sqrt{t_A}|\textbf{b}_2|)K_0(\sqrt{t_B}|\textbf{b}_q|) h_1(|\textbf{b}'_2-\textbf{b}_2+\textbf{b}_q|,t_C,t_D) \delta^2(\textbf{b}_3-\textbf{b}_2+\textbf{b}'_2) \delta^2(\textbf{b}'_3)$\\
		
		$C_{f4}$  &$\frac{1}{(2\pi)^3}K_0(\sqrt{t_A}|\textbf{b}_2|)  K_0(\sqrt{t_B}|\textbf{b}_2-\textbf{b}_3-\textbf{b}'_2+\textbf{b}_q|) K_0(\sqrt{t_C}|\textbf{b}_q-\textbf{b}_3|) K_0(\sqrt{t_D}|\textbf{b}_2-\textbf{b}_3-\textbf{b}'_2|) \delta^2(\textbf{b}_2-\textbf{b}_3-\textbf{b}'_2+\textbf{b}'_3)$\\
		
		
		$C_{g2}$  &$K_0(\sqrt{t_D}|\textbf{b}'_3|) h_2(|\textbf{b}_3+\textbf{b}'_3|,|\textbf{b}_2-\textbf{b}_3-\textbf{b}'_3|,t_A,t_B,t_C) \delta^2(\textbf{b}_3-\textbf{b}_2+\textbf{b}'_2) \delta^2(\textbf{b}_q-\textbf{b}_3)$\\
		
		$C_{g3}$  &$K_0(\sqrt{t_D}|\textbf{b}_3-\textbf{b}_q|) h_2(|\textbf{b}_q|,|\textbf{b}_2-\textbf{b}_q|,t_A,t_B,t_C) \delta^2(\textbf{b}_3-\textbf{b}_2+\textbf{b}'_2) \delta^2(\textbf{b}'_3)$\\
		
		$C_{g4}$  &$K_0(\sqrt{t_D}|\textbf{b}'_3|) h_2(|\textbf{b}_3+\textbf{b}'_3|,|\textbf{b}_2-\textbf{b}_3|,t_A,t_B,t_C) \delta^2(\textbf{b}_3-\textbf{b}_2+\textbf{b}'_2-\textbf{b}'_3) \delta^2(\textbf{b}_q-\textbf{b}_3)$\\
		
		$E_{a1}$  &$\frac{1}{(2\pi)^3}K_0(\sqrt{t_A}|\textbf{b}_3-\textbf{b}_2|)  K_0(\sqrt{t_B}|\textbf{b}'_2-\textbf{b}'_3+\textbf{b}_3|) K_0(\sqrt{t_C}|\textbf{b}_2-\textbf{b}_3-\textbf{b}'_2+\textbf{b}'_3|) K_0(\sqrt{t_D}|\textbf{b}'_3-\textbf{b}_3|) \delta^2(\textbf{b}_q)$ \\
		
		$E_{a2}$  &$\frac{1}{(2\pi)^3}K_0(\sqrt{t_A}|\textbf{b}'_3-\textbf{b}'_2|)  K_0(\sqrt{t_B}|\textbf{b}_2|) K_0(\sqrt{t_C}|\textbf{b}_3-\textbf{b}_2+\textbf{b}'_2-\textbf{b}'_3|) K_0(\sqrt{t_D}|\textbf{b}'_3-\textbf{b}_3|) \delta^2(\textbf{b}_q)$ \\
		
		$E_{a3}$  &$K_0(\sqrt{t_A}|\textbf{b}_3|) h_2(|\textbf{b}'_3-\textbf{b}_3|,|\textbf{b}_2-\textbf{b}_3+\textbf{b}'_3|,t_C,t_B,t_D) \delta^2(\textbf{b}_3-\textbf{b}_2+\textbf{b}'_2-\textbf{b}'_3) \delta^2(\textbf{b}_q)$\\
		
		$E_{a5}$  &$K_0(\sqrt{t_A}|\textbf{b}'_3|)K_0(\sqrt{t_B}|\textbf{b}_2|) h_1(|\textbf{b}_3-\textbf{b}'_3|,t_C,t_D) \delta^2(\textbf{b}_3-\textbf{b}_2+\textbf{b}'_2-\textbf{b}'_3) \delta^2(\textbf{b}_q)$\\
		
		$E_{a6}$  &$\frac{1}{(2\pi)^3}K_0(\sqrt{t_A}|\textbf{b}'_3+\textbf{b}_q|)  K_0(\sqrt{t_B}|\textbf{b}_2|) K_0(\sqrt{t_C}|\textbf{b}_q|) K_0(\sqrt{t_D}|\textbf{b}'_3-\textbf{b}_3|) \delta^2(\textbf{b}_3-\textbf{b}_2+\textbf{b}'_2-\textbf{b}'_3)$ \\
		
		$E_{a7}$  &$\frac{1}{(2\pi)^3}K_0(\sqrt{t_A}|\textbf{b}'_3-\textbf{b}_q|)  K_0(\sqrt{t_B}|\textbf{b}_2|) K_0(\sqrt{t_C}|\textbf{b}_q|) K_0(\sqrt{t_D}|\textbf{b}'_3-\textbf{b}_3|) \delta^2(\textbf{b}_3-\textbf{b}_2+\textbf{b}'_2-\textbf{b}'_3)$ \\
		
		$E_{b1}$  &$\frac{1}{(2\pi)^3}K_0(\sqrt{t_A}|\textbf{b}_3-\textbf{b}_2|)  K_0(\sqrt{t_B}|\textbf{b}'_2|) K_0(\sqrt{t_C}|\textbf{b}_2-\textbf{b}_3+\textbf{b}'_3-\textbf{b}'_2|) K_0(\sqrt{t_D}|\textbf{b}_3-\textbf{b}'_3|) \delta^2(\textbf{b}_q)$ \\
		
		$E_{b2}$  &$\frac{1}{(2\pi)^3}K_0(\sqrt{t_A}|\textbf{b}'_3-\textbf{b}'_2|)  K_0(\sqrt{t_B}|\textbf{b}_2-\textbf{b}_3+\textbf{b}'_3|) K_0(\sqrt{t_C}|\textbf{b}_3-\textbf{b}_2+\textbf{b}'_2-\textbf{b}'_3|) K_0(\sqrt{t_D}|\textbf{b}_3-\textbf{b}'_3|) \delta^2(\textbf{b}_q)$ \\
		
		$E_{b4}$  &$K_0(\sqrt{t_B}|\textbf{b}_2-\textbf{b}_3+\textbf{b}'_3|) h_2(|\textbf{b}_3-\textbf{b}'_3|,|\textbf{b}_3|,t_D,t_A,t_C) \delta^2(\textbf{b}_3-\textbf{b}_2+\textbf{b}'_2-\textbf{b}'_3) \delta^2(\textbf{b}_q)$\\
		
		$E_{b6}$  &$\frac{1}{(2\pi)^3}K_0(\sqrt{t_A}|\textbf{b}'_3+\textbf{b}_q|)  K_0(\sqrt{t_B}|\textbf{b}_2-\textbf{b}_3+\textbf{b}'_3|) K_0(\sqrt{t_C}|\textbf{b}_q|) K_0(\sqrt{t_D}|\textbf{b}_3-\textbf{b}'_3|) \delta^2(\textbf{b}_3-\textbf{b}_2+\textbf{b}'_2-\textbf{b}'_3)$ \\
		
		$E_{b7}$  &$\frac{1}{(2\pi)^3}K_0(\sqrt{t_A}|\textbf{b}_3-\textbf{b}_q|)  K_0(\sqrt{t_B}|\textbf{b}_2-\textbf{b}_3+\textbf{b}'_3|) K_0(\sqrt{t_C}|\textbf{b}_q|) K_0(\sqrt{t_D}|\textbf{b}_3-\textbf{b}'_3|) \delta^2(\textbf{b}_3-\textbf{b}_2+\textbf{b}'_2-\textbf{b}'_3)$ \\
		
		$E_{c1}$  &$\frac{1}{(2\pi)^3}K_0(\sqrt{t_A}|\textbf{b}_3-\textbf{b}_2|)  K_0(\sqrt{t_B}|\textbf{b}'_2+\textbf{b}_q|) K_0(\sqrt{t_C}|\textbf{b}_2-\textbf{b}'_2|) K_0(\sqrt{t_D}|\textbf{b}_q|) \delta^2(\textbf{b}_3-\textbf{b}'_3)$ \\
		
		$E_{c2}$  &$\frac{1}{(2\pi)^3}K_0(\sqrt{t_A}|\textbf{b}_3-\textbf{b}'_2|)  K_0(\sqrt{t_B}|\textbf{b}_2+\textbf{b}_q|) K_0(\sqrt{t_C}|\textbf{b}'_2-\textbf{b}_2|) K_0(\sqrt{t_D}|\textbf{b}_q|) \delta^2(\textbf{b}'_3-\textbf{b}_3)$ \\
		
		$E_{c5}$  &$K_0(\sqrt{t_A}|\textbf{b}_3+\textbf{b}_q|) h_2(|\textbf{b}_q|,|\textbf{b}_2+\textbf{b}_q|,t_D,t_B,t_C) \delta^2(\textbf{b}'_2-\textbf{b}_2) \delta^2(\textbf{b}'_3-\textbf{b}_3)$\\
		
		$E_{c7}$  &$h_3(|\textbf{b}_3-\textbf{b}_q|,|\textbf{b}_2|,|\textbf{b}_q|,t_A,t_B,t_C,t_D) \delta^2(\textbf{b}'_2-\textbf{b}_2) \delta^2(\textbf{b}'_3-\textbf{b}_3)$\\
		
		$E_{d1}$  &$\frac{1}{(2\pi)^3}K_0(\sqrt{t_A}|\textbf{b}_3-\textbf{b}_2|)  K_0(\sqrt{t_B}|\textbf{b}'_2-\textbf{b}_q|) K_0(\sqrt{t_C}|\textbf{b}_2-\textbf{b}'_2|) K_0(\sqrt{t_D}|\textbf{b}_q|) \delta^2(\textbf{b}'_3-\textbf{b}_3)$ \\
		
		$E_{d2}$  &$\frac{1}{(2\pi)^3}K_0(\sqrt{t_A}|\textbf{b}_3-\textbf{b}'_2|)  K_0(\sqrt{t_B}|\textbf{b}_2-\textbf{b}_q|) K_0(\sqrt{t_C}|\textbf{b}'_2-\textbf{b}_2|) K_0(\sqrt{t_D}|\textbf{b}_q|) \delta^2(\textbf{b}'_3-\textbf{b}_3)$ \\
		
		$E_{d6}$  &$K_0(\sqrt{t_B}|\textbf{b}_2+\textbf{b}_q|) h_2(|\textbf{b}_q|,|\textbf{b}_3+\textbf{b}_q|,t_D,t_A,t_C) \delta^2(\textbf{b}'_2-\textbf{b}_2) \delta^2(\textbf{b}'_3-\textbf{b}_3)$\\
		
		\hline\hline
	\end{tabular}
\end{table}

\begin{table} [H]	
	\scriptsize
	\centering
	\caption{The virtualities of the internal propagators $t_{A,B,C,D}$  for the $W$-emission and $W$-exchange diagrams.}
	\newcommand{\tabincell}[2]{\begin{tabular}{@{}#1@{}}#2\end{tabular}}
	\label{tab:ttt}
	\begin{tabular}[t]{lcccc}
		\hline\hline\
		$R_{ij}$ &  $\frac{t_A}{M^2}$   &$\frac{t_B}{M^2}$  &$\frac{t_C}{M^2}$ &$\frac{t_D}{M^2}$ \\ \hline
		$C_{a1}$ &$f^+x_1y$ &$f^+(f^--1)x_2x_2'$ &$f^+(f^--y)(x_2-1)+r_c^2$ &$f^+(f^-(x_2'-1)-x_2'+y)+r_c^2$ \\
		
		$C_{a2}$ &$f^+x_1y$ &$f^+(f^--1)x_2x_2'$ &$f^+(x_1-1)(f^-(1-x_2')+x_2')+r_c^2$ &$f^+(f^-(x_2'-1)-x_2'+y)+r_c^2$ \\	
			
		$C_{a3}$ &$f^+x_1y$ &$f^+(f^--1)x_2x_2'$ &$f^+x_1((f^--1)x_3'+y)$ &$f^+(x_1-1)(f^-(1-x_2')+x_2')+r_c^2$ \\
		
		$C_{a4}$ &$f^+x_1y$ &$f^+ \left(x_1+x_2\right) \left(\left(f^--1\right) x'_2+y\right)$   &$f^+y(x_1+x_2)$ &$r_c^2+f^+ \left(f^- \left(x'_2-1\right)-x'_2+y\right)$ \\	
			
		$C_{a5}$ &$f^+x_1y$ &$f^+(x_1+x_2)((f^--1)x_2'+y)$ &$f^+x_1((f^--1)x_2'+y)$ &$f^+(f^-(x_2'-1)-x_2'+y)+r_c^2$ \\
		
		$C_{a6}$ &$f^+x_1y$ &$f^+(f^--1)x_2x_2'$ &$f^+x_1$ &$f^+(x_1-1)(f^-(1-x_2')+x_2')+r_c^2$ \\	
			
		$C_{a7}$ &$f^+x_1y$ &$f^+(f^--1)x_2x_2'$ &$(f^+(x_1-1)+1)((f^--1)x_1'+y)+r_c^2$ &$f^+(x_1-1)(f^-(1-x_2')+x_2')+r_c^2$ \\
		
		$C_{b1}$ &$f^+x_1y$ &$f^+(f^--1)x_2x_2'$ &$f^+(x_2-1)(f^--y)+r_c^2$ &$f^+(f^--1)x_2(x_2'+x_3')$ \\	
			
		$C_{b2}$ &$f^+x_1y$ &$f^+(f^--1)x_2x_2'$ &$f^+(x_1+x_2)((f^--1)(x_2'+x_3')+y)$ &$f^+(f^--1)x_2(x_2'+x_3')$ \\
		
		$C_{b3}$ &$f^+x_1y$ &$f^+(f^--1)x_2x_2'$ &$f^+x_1((f^--1)x_3'+y)$ &$f^+(x_1+x_2)((f^--1)(x_2'+x_3')+y)$ \\	
			
		$C_{b4}$ &$f^+x_1y$ &$f^+(x_1+x_2)((f^--1)x_2'+y)$ &$f^+y(x_1+x_2)$ &$f^+(x_1+x_2)((f^--1)(x_2'+x_3')+y)$ \\
		
		$C_{b5}$ &$f^+x_1y$ &$f^+(x_1+x_2)((f^--1)x_2'+y)$ &$f^+x_1((f^--1)x_2'+y)$ &$f^+(x_1+x_2)((f^--1)(x_2'+x_3')+y)$ \\	
			
		$C_{b6}$ &$f^+x_1y$  &$\left(f^--1\right) f^+ x_2 x'_2$ &$f^+x_1$ &$\left(f^--1\right) f^+ x_2 \left(x'_2+x'_3\right)$ \\
		
		$C_{b7}$ &$f^+x_1y$ &$\left(f^--1\right) f^+ x_2 x'_2$ &$(f^+(x_1-1)+1)((f^--1)x_1'+y)+r_c^2$ &$\left(f^--1\right) f^+ x_2 \left(x'_2+x'_3\right)$ \\	
			
		$C_{c1}$ &$f^+x_1y$ &$\left(f^--1\right) f^+ x_2 x'_2$ &$f^+(x_2-1)(f^--y)+r_c^2$ &\tabincell{l}{$-f^-(x_3'-1)(f^+(x_2-1)+1)$\\$+f^+(x_2-1)(x_3'-1)+r_c^2+x_3'-1$} \\
		
		$C_{c2}$ &$f^+x_1y$ &$\left(f^--1\right) f^+ x_2 x'_2$ &$f^+x_1((f^--1)x_3'+y)$ &\tabincell{l}{$-f^-(x_3'-1)(f^+(x_2-1)+1)$\\$+f^+(x_2-1)(x_3'-1)+r_c^2+x_3'-1$} \\	
			
		$C_{c3}$ &$f^+x_1y$ &$f^+(x_1+x_2)((f^--1)x_2'+y)$ &$f^+y(x_1+x_2)$ &$(f^+x_3-1)[f^-(x_3'-1)-(x_3'+y-1)]+r_c^2$ \\
		
		$C_{c4}$ &$f^+x_1y$ &$f^+(x_1+x_2)((f^--1)x_2'+y)$ &$f^+x_1(y+(f^--1)x_2')$ &$(f^+x_3-1)[f^-(x_3'-1)-(x_3'+y-1)]+r_c^2$ \\
			
		$C_{c5}$ &$f^+x_1y$ &$(f^--1)f^+x_2x_2'$ &$f^+x_1$ &\tabincell{l}{$-f^-(x_3'-1)(f^+(x_2-1)+1)$\\$+f^+(x_2-1)(x_3'-1)+r_c^2+x_3'-1$} \\
		
		$C_{c6}$ &$f^+x_1y$ &$(f^--1)f^+x_2x_2'$ &\tabincell{l}{$f^-(x_3'-1)(f^+x_3-1)$\\$-(f^+x_3-1)(x_3'+y-1)+r_c^2$} &\tabincell{l}{$-f^-(x_3'-1)(f^+(x_2-1)+1)$\\$+f^+(x_2-1)(x_3'-1)+r_c^2+x_3'-1$} \\	
			
		$C_{c7}$ &$f^+x_1y$ &$(f^--1)f^+x_2x_2'$ &$(f^+(x_1-1)+1)((f^--1)x_1'+y)+r_c^2$ &$(f^+x_3-1)[f^-(x_3'-1)-(x_3'+y-1)]+r_c^2$ \\
		
		$C_{d1}$ &$f^+x_1y$ &$(f^--1)f^+x_2x_2'$ &$f^+(x_2-1)(f^--y)+r_c^2$ &$f^+x_2((f^--1)x_2'-y+1)-f^-x_2'+x_2'+y$ \\	
			
		$C_{d2}$ &$f^+x_1y$ &$(f^--1)f^+x_2x_2'$ &$f^+x_1((f^--1)x_3'+y)$ &$f^+x_2((f^--1)x_2'-y+1)-f^-x_2'+x_2'+y$ \\
		
		$C_{d3}$ &$f^+x_1y$ &$f^+(x_1+x_2)((f^--1)x_2'+y)$ &$f^+y(x_1+x_2)$ &$-f^+(x_3-1)((f^--1)x_2'+1)-f^-x_2'+x_2'$ \\	
			
		$C_{d4}$ &$f^+x_1y$ &$f^+(x_1+x_2)((f^--1)x_2'+y)$ &$f^+x_1(y+(f^--1)x_2')$ &$-f^+(x_3-1)((f^--1)x_2'+1)-f^-x_2'+x_2'$ \\
		
		$C_{d5}$ &$f^+x_1y$ &$(f^--1)f^+x_2x_2'$ &$f^+x_1$ &$-f^+(x_3-1)((f^--1)x_2'+1)-f^-x_2'+x_2'$ \\	
			
		$C_{d6}$ &$f^+x_1y$ &$(f^--1)f^+x_2x_2'$ &$f^+((f^--1)x_2'+1)(x_1+x_2)-f^-x_2'+x_2'$ &$f^+x_2((f^--1)x_2'-y+1)-f^-x_2'+x_2'+y$ \\
		
		$C_{d7}$ &$f^+x_1y$ &$(f^--1)f^+x_2x_2'$ &$(f^+(x_1-1)+1)((f^--1)x_1'+y)+r_c^2$ &$f^+x_2((f^--1)x_2'-y+1)-f^-x_2'+x_2'+y$ \\	
			
		$C_{e1}$ &$f^+(f^--1)x_2x_2'$ &$f^+(1-x_3)((f^--1)x_2'+y)$ &$f^+x_2((f^--1)x_2'+y)$ &$f^+(f^-(x_2'-1)-x_2'+y)+r_c^2$ \\
		
		$C_{e2}$ &$f^+(f^--1)x_2x_2'$ &$f^+(1-x_3)((f^--1)x_2'+y)$ &$f^+x_2((f^--1)x_2'+y)$ &$f^+(x_1+x_2)((f^--1)(x_2'+x_3')+y)$ \\	
			
		$C_{e3}$ &$f^+(f^--1)x_2x_2'$ &$f^+(1-x_3)((f^--1)x_2'+y)$ &$f^+x_2((f^--1)x_2'+y)$ &$f^+((f^--1)x_2'+1)(x_1+x_2)-f^-x_2p+x_2'$ \\
		
		$C_{e4}$ &$f^+(f^--1)x_2x_2'$ &$f^+(1-x_3)((f^--1)x_2'+y)$ &$f^+x_2((f^--1)x_2'+y)$ &$(f^+x_3-1)[f^-(x_3'-1)-(x_3'+y-1)]+r_c^2$ \\
				
		$C_{f1}$ &$f^+(f^--1)x_2x_2'$ &$f^+(1-x_3)((f^--1)x_2'+y)$ &$(f^--1)f^+x_2'(x_1+x_2)$ &$r_c^2+f^+ \left(f^- \left(x'_2-1\right)-x'_2+y\right)$ \\
		
		$C_{f2}$ &$f^+(f^--1)x_2x_2'$ &$f^+(1-x_3)((f^--1)x_2'+y)$ &$(f^--1)f^+x_2'(x_1+x_2)$ &$f^+ \left(-\left(x_3-1\right)\right) \left(\left(f^--1\right)
   \left(x'_2+x'_3\right)+y\right)$ \\	
			
		$C_{f3}$ &$f^+(f^--1)x_2x_2'$ &$f^+(1-x_3)((f^--1)x_2'+y)$ &$(f^--1)f^+x_2'(x_1+x_2)$ &$-f^+(x_3-1)((f^--1)x_2'+1)-f^-x_2'+x_2'$ \\
		
		$C_{f4}$ &$f^+(f^--1)x_2x_2'$ &$f^+(1-x_3)((f^--1)x_2'+y)$ &$(f^--1)f^+x_2'(x_1+x_2)$ &$(f^+x_3-1)[f^-(x_3'-1)-(x_3'+y-1)]+r_c^2$ \\
			
		$C_{g2}$ &$f^+x_1y$ &$(f^--1)f^+x_2x_2'$ &$f^+(x_1+x_2)((f^--1)x_2'+y)$ &$f^+(x_1+x_2)((f^--1)(x_2'+x_3')+y)$ \\	
			
		$C_{g3}$ &$f^+x_1y$ &$(f^--1)f^+x_2x_2'$ &$f^+(x_1+x_2)((f^--1)x_2'+y)$ &$f^+((f^--1)x_2'+1)(x_1+x_2)-f^-x_2'+x_2'$ \\
		
		$C_{g4}$ &$f^+x_1y$ &$(f^--1)f^+x_2x_2'$ &$f^+(x_1+x_2)((f^--1)x_2'+y)$ &$(f^+x_3-1)[f^-(x_3'-1)-(x_3'+y-1)]+r_c^2$ \\
		
		$E_{a1}$ &$(f^--1)f^+x_3x_3'$ &$(f^--1)f^+(1-x_1)(1-x_1')$ &$(f^--1)f^+x_3'(1-x_1)$ &$r_c^2+f^+(x_1'-1-f^-x_1')$ \\	
		
		$E_{a2}$ &$(f^--1)f^+x_3x_3'$ &$(f^--1)f^+(1-x_1)(1-x_1')$ &$(f^--1)f^+x_3(1-x_1')$ &$r_c^2+f^+(x_1'-1-f^-x_1')$ \\
		
		$E_{a3}$ &$(f^--1)f^+x_3x_3'$ &$(f^--1)f^+x_2x_2'$ &$r_c^2+f^+(x_2-1)(f^-+x_3'-f^-x_3')$ &$r_c^2+f^+(x_1'-1-f^-x_1')$ \\
		
		$E_{a5}$ &$(f^--1)f^+x_3x_3'$ &$(f^--1)f^+x_2x_2'$&\tabincell{l}{$r_c^2-1+x_2'-f^-(x_2'-1)$\\$(1+f^+(x_3-1))+f^+(x_2'-1)(x_3-1)$} &$r_c^2-f^+ \left(x_3-1\right) \left(f^-
   \left(x'_2-1\right)-x'_2\right)$ \\
		
		$E_{a6}$ &$(f^--1)f^+x_3x_3'$ &$(f^--1)f^+x_2x_2'$ &$f^+ x_3 \left(\left(f^--1\right) x'_3+y\right)$ &$r_c^2-f^+ \left(x_3-1\right) \left(f^-
   \left(x'_2-1\right)-x'_2\right)$ \\
		
		$E_{a7}$ &$(f^--1)f^+x_3x_3'$ &$(f^--1)f^+x_2x_2'$ &$f^+ x_3 \left(\left(f^--1\right) x'_3-y+1\right)+x'_3(1-f^-)+y$ &$r_c^2-f^+ \left(x_3-1\right) \left(f^-
   \left(x'_2-1\right)-x'_2\right)$ \\
		
		$E_{b1}$ &$(f^--1)f^+x_3x_3'$ &$(f^--1)f^+(1-x_1)(1-x_1')$ &$(f^--1)f^+(1-x_1)x_3'$ &$f^--1+r_c^2+f^+x_1-f^-f^+x_1$ \\
		
		$E_{b2}$ &$(f^--1)f^+x_3x_3'$ &$(f^--1)f^+(1-x_1)(1-x_1')$ &$(f^--1)f^+x_3(1-x_1')$ &$r_c^2+f^--1+f^+x_1-f^-f^+x_1$ \\
		
		$E_{b4}$ &$(f^--1)f^+x_3x_3'$ &$(f^--1)f^+x_2x_2'$ &$r_c^2+f^--1+f^+x_1-f^-f^+x_1$ & \tabincell{l}{$r_c^2-1-f^-(1+f^+(x_2-1))$\\$(x_3'-1)+f^+(x_2-1)(x_3'-1)+x_3'$}\\
		
		$E_{b6}$ &$(f^--1)f^+x_3x_3'$ &$(f^--1)f^+x_2x_2'$ &$f^+x_3((f^--1)x_3'+y)$ &\tabincell{l}{$r_c^2-1-f^-(1+f^+(x_2-1))$\\$(x_3'-1)+f^+(x_2-1)(x_3'-1)+x_3'$}\\
		
		$E_{b7}$ &$(f^--1)f^+x_3x_3'$ &$(f^--1)f^+x_2x_2'$ &$x_3'-f^-x_3'+f^+x_3(1+(f^--1)x_3'-y)+y$ &\tabincell{l}{$r_c^2-1-f^-(1+f^+(x_2-1))$\\$(x_3'-1)+f^+(x_2-1)(x_3'-1)+x_3'$}\\
		
		$E_{c1}$ &$(f^--1)f^+x_3x_3'$ &$(f^--1)f^+(1-x_1)(1-x_1')$ &$f^+x_3'(f^--1+x_1-f^-x_1)$ &$f^+(1-x_1)((f^--1)(1-x_1')+y)$ \\
		
		$E_{c2}$ &$(f^--1)f^+x_3x_3'$ &$(f^--1)f^+(1-x_1)(1-x_1')$ &$(f^--1)f^+x_3(1-x_1')$ &$f^+(1-x_1)((f^--1)(1-x_1')+y)$ \\
		
		$E_{c5}$ &$(f^--1)f^+x_3x_3'$ &$(f^--1)f^+x_2x_2'$ &$f^+x_3((f^--1)x_3'+y)$ &$f^+(1-x_1)((f^--1)(1-x_1')+y)$ \\
		
		$E_{c7}$ &$(f^--1)f^+x_3x_3'$ &$(f^--1)f^+x_2x_2'$ &$f^+x_3(1+(f^--1)x_3'-y)+x_3'-f^-x_3'+y$ &$f^+x_2((f^--1)x_2'+y)$ \\
		
		$E_{d1}$ &$(f^--1)f^+x_3x_3'$ &$(f^--1)f^+(1-x_1)(1-x_1')$ &$f^+(f^--1+x_1-f^-x_1)x_3'$ &\tabincell{l}{$y-\left(f^--1\right) \left(x'_2+x'_3\right)+f^+
   \left(x_2+x_3\right)$\\$ \left(\left(f^--1\right)
   x'_2+\left(f^--1\right) x'_3-y+1\right)$}\\
		
		$E_{d2}$ &$(f^--1)f^+x_3x_3'$ &$(f^--1)f^+(1-x_1)(1-x_1')$ &$(f^--1)f^+x_3(1-x_1')$ &\tabincell{l}{$y-\left(f^--1\right) \left(x'_2+x'_3\right)+f^+
   \left(x_2+x_3\right)$\\$ \left(\left(f^--1\right)
   x'_2+\left(f^--1\right) x'_3-y+1\right)$}\\
		
		$E_{d6}$ &$(f^--1)f^+x_3x_3'$ &$(f^--1)f^+(1-x_1)(1-x_1')$  &\tabincell{l}{$y-\left(f^--1\right) \left(x'_2+x'_3\right)+f^+
   \left(x_2+x_3\right)$\\$ \left(\left(f^--1\right)
   x'_2+\left(f^--1\right) x'_3-y+1\right)$} &$f^+x_2(1+(f^--1)x_2'-y)+x_2'-f^-x_2'+y$\\
		
		\hline\hline
	\end{tabular}
\end{table}

\end{appendix}

\end{document}